\documentclass[twocolumn]{aastex631}

\usepackage{amssymb, amsmath}
\usepackage{color}
\usepackage{float}

\usepackage{comment} 
\accepted{for publication in ApJ on 6 February, 2024}
\begin{document}

\title{
EMPRESS. XIII.\\
Chemical Enrichments of Young Galaxies Near and Far at $z\sim 0$ and $4-10$:\\
Fe/O, Ar/O, S/O, and N/O Measurements with Chemical Evolution Model Comparisons}

\author{Kuria Watanabe}
\affiliation{Department of Astronomical Science, SOKENDAI (The Graduate University for Advanced Studies), \\ 2-21-1 Osawa, Mitaka, Tokyo, 181-8588, Japan}
\affiliation{National Astronomical Observatory of Japan, 2-21-1 Osawa, Mitaka, Tokyo, 181-8588, Japan}

\author{Masami Ouchi}
\affiliation{National Astronomical Observatory of Japan, 2-21-1 Osawa, Mitaka, Tokyo, 181-8588, Japan}
\affiliation{Institute for Cosmic Ray Research, The University of Tokyo, 5-1-5 Kashiwa-no-Ha, Kashiwa, Chiba, 277-8582, Japan}
\affiliation{Department of Astronomical Science, SOKENDAI (The Graduate University for Advanced Studies), \\ 2-21-1 Osawa, Mitaka, Tokyo, 181-8588, Japan}
\affiliation{Kavli Institute for the Physics and Mathematics of the Universe (WPI), The University of Tokyo, Kashiwa, Chiba 277-8583, Japan}

\author{Kimihiko Nakajima}
\affiliation{National Astronomical Observatory of Japan, 2-21-1 Osawa, Mitaka, Tokyo, 181-8588, Japan}

\author{Yuki Isobe}
\affiliation{Institute for Cosmic Ray Research, The University of Tokyo, 5-1-5 Kashiwa-no-Ha, Kashiwa, Chiba, 277-8582, Japan}
\affiliation{Department of Physics, Graduate School of Science, The University of Tokyo, 7-3-1 Hongo, Bunkyo, Tokyo 113-0033, Japan}

\author{Nozomu Tominaga}
\affiliation{National Astronomical Observatory of Japan, 2-21-1 Osawa, Mitaka, Tokyo, 181-8588, Japan}
\affiliation{Department of Physics, Faculty of Science and Engineering, Konan University, 8-9-1 Okamoto, Kobe, Hyogo 658-8501, Japan}
\affiliation{Department of Astronomical Science, SOKENDAI (The Graduate University for Advanced Studies), \\ 2-21-1 Osawa, Mitaka, Tokyo, 181-8588, Japan}

\author{Akihiro Suzuki}
\affiliation{Research Center for the Early Universe, Graduate School of Science, The University of Tokyo, \\7-3-1 Hongo, Bunkyo, Tokyo 113-0033, Japan}

\author{Miho N. Ishigaki}
\affiliation{National Astronomical Observatory of Japan, 2-21-1 Osawa, Mitaka, Tokyo, 181-8588, Japan}

\author{Ken'ichi Nomoto}
\affiliation{Kavli Institute for the Physics and Mathematics of the Universe (WPI), The University of Tokyo, Kashiwa, Chiba 277-8583, Japan}

\author{Koh Takahashi}
\affiliation{National Astronomical Observatory of Japan, 2-21-1 Osawa, Mitaka, Tokyo, 181-8588, Japan}

\author{Yuichi Harikane}
\affiliation{Institute for Cosmic Ray Research, The University of Tokyo, 5-1-5 Kashiwa-no-Ha, Kashiwa, Chiba, 277-8582, Japan}

\author{Shun Hatano}
\affiliation{Department of Astronomical Science, SOKENDAI (The Graduate University for Advanced Studies), \\ 2-21-1 Osawa, Mitaka, Tokyo, 181-8588, Japan}
\affiliation{National Astronomical Observatory of Japan, 2-21-1 Osawa, Mitaka, Tokyo, 181-8588, Japan}

\author{Haruka Kusakabe}
\affiliation{Observatoire de Gen`eve, Universit´e de Gen`eve, 51 Chemin de P´egase, 1290 Versoix, Switzerland}
\affiliation{National Astronomical Observatory of Japan, 2-21-1 Osawa, Mitaka, Tokyo, 181-8588, Japan}

\author{Takashi J. Moriya}
\affiliation{National Astronomical Observatory of Japan, 2-21-1 Osawa, Mitaka, Tokyo, 181-8588, Japan}
\affiliation{School of Physics and Astronomy, Faculty of Science, Monash University, Clayton, Victoria 3800, Australia}

\author{Moka Nishigaki}
\affiliation{Department of Astronomical Science, SOKENDAI (The Graduate University for Advanced Studies), \\ 2-21-1 Osawa, Mitaka, Tokyo, 181-8588, Japan}
\affiliation{National Astronomical Observatory of Japan, 2-21-1 Osawa, Mitaka, Tokyo, 181-8588, Japan}

\author{Yoshiaki Ono}
\affiliation{Institute for Cosmic Ray Research, The University of Tokyo, 5-1-5 Kashiwa-no-Ha, Kashiwa, Chiba, 277-8582, Japan}

\author{Masato Onodera}
\affiliation{National Astronomical Observatory of Japan, 2-21-1 Osawa, Mitaka, Tokyo, 181-8588, Japan}
\affiliation{Subaru Telescope, National Astronomical Observatory of Japan, National Institutes of Natural Sciences (NINS),\\ 650 North A’ohoku Place, Hilo, HI 96720, USA}

\author{Yuma Sugahara}
\affiliation{Waseda Research Institute for Science and Engineering, Faculty of Science and Engineering, Waseda University, 3-4-1, Okubo, Shinjuku, Tokyo 169-8555, Japan}
\affiliation{National Astronomical Observatory of Japan, 2-21-1 Osawa, Mitaka, Tokyo, 181-8588, Japan}

\begin{abstract}
We present gas-phase elemental abundance ratios of 7 local extremely metal-poor galaxies (EMPGs) including our new Keck/LRIS spectroscopy determinations together with 33 JWST $z\sim 4-10$ star-forming galaxies in the literature, and compare chemical evolution models. We develop chemical evolution models with the yields of core-collapse supernovae (CCSNe), Type Ia supernovae, hypernovae (HNe), and pair-instability supernovae (PISNe), and compare the EMPGs and high-$z$ galaxies in conjunction with dust depletion contributions. We find that high Fe/O values of EMPGs can (cannot) be explained by PISN metal enrichments (CCSN/HN enrichments even with the mixing-and-fallback mechanism enhancing iron abundance), while that the observed Ar/O and S/O values are much smaller than the predictions of the PISN models. The abundance ratios of the EMPGs can be explained by the combination of Type Ia SNe and CCSNe/HNe whose inner layers of argon and sulfur mostly fallback, which are comparable with Sculptor stellar chemical abundance distribution, suggesting that early chemical enrichment is taken place in the EMPGs. 
Comparing our chemical evolution models with the star-forming galaxies at $z\sim 4-10$, we find that the Ar/O and S/O ratios of the high-$z$ galaxies are comparable with those of the CCSN/HN models, while majority of the high-$z$ galaxies do not have constraints good enough to rule out contributions from PISNe.
The high N/O ratio recently reported in GN-z11 cannot be explained even by rotating PISNe,
but could be reproduced by the winds of rotating Wolf Rayet stars that end up as a direct collapse.

\end{abstract}


\section{Introduction} \label{sec:intro}
Chemical properties of young galaxies are important to understand
the chemical evolution in galaxy formation.
Numerical simulations are conducted to reproduce galaxies at the early formation phase
(\citealt{2022MNRAS.509.4037Y}; \citealt{2012ApJ...745...50W}).
\cite{2012ApJ...745...50W} find that early galaxies with the stellar age of 300 Myr have low metallicities of $0.1~\% -1~\%$ solar abundance and low stellar masses of $10^4 - 10^9 M_\odot$ at high redshift of $z\geq 7$.
%
There are observational studies for early galaxies at $z\geq 7$ (e.g., \citealt{2018MNRAS.479.1180M}; \citealt{2016ARA&A..54..761S}). However, early galaxies at high-$z$ are too faint to be detected because of their low stellar masses.
There are efforts of observations on galaxies at the early formation phase in the low-redshift universe
(e.g., \citealt{2019ApJ...874...93B}; \citealt{2014MNRAS.445.3200S}),
while even with James Webb Space Telescope (JWST) it is difficult to detect early galaxies with low masses such as $M_* \leq 10^6 M_\odot$ at $z\geq 2$ \citep{2020ApJ...893...60K}.

Although early galaxies are not well observationally investigated yet, one can study the early phase of galaxy formation in the local universe with dwarf galaxies (e.g., \citealt{2021ApJ...918...54I},\citealt{2021A&A...646A.138I}). 
It should be noted that galaxy formation in the early universe may be different from the one in the local universe, studies of galaxy formation in the local universe serve the first step for understanding the galaxy formation at high redshift.
 
Various studies show the presence of extremely metal-poor galaxies (EMPGs) such as SBS 0335-052 \citep{2009A&A...503...61I}, AGC198691 \citep{2016ApJ...822..108H}, J1234+3901 \citep{2019MNRAS.483.5491I}, and I Zw 18 \citep{1998ApJ...497..227I}.  
EMPGs are defined as galaxies with less than 10 \% solar oxygen abundance.
Recently, \cite{2020ApJ...898..142K} have launched a project named ``Extremely Metal-Poor Representative Explored by the Subaru Survey (EMPRESS)".
The EMPRESS project searches for EMPGs by the machine learning methods with the Subaru and Sloan Digital Sky Survey data \citep{2021ApJ...913...22K,2023arXiv230203158N}, and studies their physical properties, for examples, morphologies \citep{2021ApJ...918...54I}, outflows \citep{2022ApJ...929..134X}, strong high-ionization lines \citep{2022ApJ...930...37U}, and He abundance \citep{2022ApJ...941..167M} of EMPGs.
One of the notable indications from the series of the EMPRESS work is a high Fe/O ratio ([Fe/O]$\sim$0) in EMPGs despite the low metallicities \citep{2021ApJ...913...22K}.
Although the primary driver of iron enrichment in galaxies is generally thought to be Type Ia supernovae (SNe), EMPGs would be too young to be enriched by Type Ia SNe due to a typically long ($\sim 10^9$ yr) delay time to happen. \cite{2022ApJ...925..111I} examine other scenarios to reproduce high Fe/O in young galaxies like EMPGs, interestingly suggesting that massive-star explosions such as hypernovae (HNe) and/or pair-instability SNe (PISNe) can explain the iron-rich and oxygen-poor properties.
\cite{2022ApJ...925..111I} show EMPGs' Fe/O as high as those enriched by PISNe or bright HNe with the models made in the same manner as \cite{2018ApJ...852..101S} with core-collapse SNe (CCSNe), HNe, and PISNe yields of \cite{2008ApJ...673.1014U}, \cite{N13papaer}, and \cite{2018ApJ...857..111T}, respectively.
CCSN is an explosion that occurs at the end of the evolution of a massive star with a mass greater than $8~M_\odot$ while HN has higher explosion energy than that of CCSN and ejects more iron.
Very massive stars ($> 140~M_\odot$) cause PISNe and have no compact remnants \citep{2018ApJ...857..111T}. 
However, a single piece of observational evidence, the high Fe/O, is not strong enough to conclude that EMPGs are mainly enriched by PISNe or bright HNe.
Moreover, one should distinguish between the contributions from PISNe and HNe because the Fe/O values given by ejecta of PISNe and HNe are comparable (\citealt{2022ApJ...925..111I}, \citealt{2002ApJ...578..855U}).
Because S/O and Ar/O values are different between ejecta of PISNe and HNe, one can distinguish the origin of the abundant Fe with S/O and Ar/O.
There should also remain signatures of PISNe or bright HNe in metal-poor stars of the present-day Milky Way (MW) and local dwarf galaxies, if PISNe or bright HNe took place at the early phase of galaxy formation. One should investigate abundance ratios of metal-poor stars in the local universe.
%

While metal-poor galaxies and stars in the local universe are important, studies of high-$z$ galaxies are also key to understanding abundance ratios of young galaxies. We should discuss abundance ratios in galaxies both at $z\sim 0$ and high redshift towards the early epoch of galaxy formation at $z\sim 10$.

%
Although it is difficult to investigate abundance ratios of high-$z$ galaxies whose observational signatures are too weak to detect, gravitational lensing magnifications allow us to detect such weak signatures in high-$z$ galaxies with JWST.
Recent studies for early chemical enrichment of high-$z$ galaxies proceed very rapidly with observational data obtained with JWST.
The diagnostic optical emission lines such as [{\sc O iii}]$\lambda \lambda$5007,4959, [{\sc O ii}]$\lambda$3727, and hydrogen Balmer lines have now been identified in galaxies at high redshift up to $z \sim 4-10$, suggesting rapid chemical enrichment in galaxies with measurements of (O/H) (\citealt{2023arXiv230408516C,2022A&A...665L...4S,2023ApJ...945...35T,2023ApJ...942L..14R,2023MNRAS.525.2087B,2023arXiv230112825N}).
\cite{2022ApJ...940L..23A} report the abundance ratios such as Ne/O and C/O of galaxies at $z \geq 7$ that are observed with JWST/NIRSpec.
Although \cite{2022ApJ...940L..23A} detect [Fe{\sc iii}]$\lambda$2465 and [Fe{\sc ii}]$\lambda$4360 lines for one of these high-$z$ galaxies, an Fe/O ratio is not determined due to the low S/N ratios of these emission lines.

%

Elemental abundance ratios in galaxies at high redshift are also drawing attention.
In particular, \cite{2023arXiv230207256B} identify emission lines of a galaxy GN-z11 at $z \geq 10$ such as [Ne {\sc iii}], [{\sc N iii}], and [{\sc O ii}] with JWST/NIRSpec data,
and claim that the galaxy at $z=10.6$ has an extremely high N/O larger than the solar abundance \citep{2023arXiv230210142C}.
The abundance ratios with high-$z$ galaxies are important to understand the chemical enrichment driven by galaxies in the early universe. 

This paper is the XIIIth paper of the EMPRESS.
In this paper, we present spectroscopic observations for EMPGs with Keck Telescope, and discuss the abundance ratios of EMPGs with the chemical evolution models.
Our observations and data reduction methods are described in Section \ref{sec:obs}.
In Section \ref{sec:sample}, we explain our sample and data analysis.
In Section \ref{sec:model}, we develop chemical evolution models of galaxies.
We present our results, and discuss the abundance ratios of the EMPGs by
comparisons with the chemical evolution models in Section \ref{sec:results}.
In Section \ref{sec:summary}, we summarize our study.
Throughout this paper, we assume a solar metallicity $Z_\odot$ as $12+\log({\rm O/H}) = 8.69$, and use the solar abundance ratios of $\log({\rm Fe/O})=-1.23$, $\log({\rm Ar/O})=-2.31$, $\log({\rm S/O})=-1.57$, $\log({\rm Ne/O})=-0.63$, and $\log({\rm N/O})=-0.86$, respectively \citep{Asplaud}.
Abundance ratios are defined by those normalized by the solar abundance ratios,
\begin{equation}
    \mathrm{[A/B]} = \log_{10} \left( \frac{N_A / N_{A,\odot}}{N_B / N_{B,\odot}} \right),
\end{equation}
where $N_A$ and $N_B$ are the numbers of the element A and B, respectively. The variables of $N_{A, \odot}$ and $N_{B,\odot}$ indicate the solar abundances.

\section{Observations and Data Reduction} \label{sec:obs}
\subsection{Enlarging the EMPG Sample}
This study needs EMPGs with measurements of Fe/O and the other various abundance ratios, Ar/O, S/O, Ne/O, and N/O. Because in the literature we find only 11 EMPGs whose Fe/O values can be determined with [Fe {\sc iii}]$\lambda$4658 emission, to increase the number of EMPGs we conduct deep spectroscopy for EMPGs with the Keck/the Low-Resolution Imaging Spectrometer (LRIS) spectrograph. We select three EMPGs, SBS-0335-052E \citep{2009A&A...503...61I}, J2314+0154 \citep{2020ApJ...898..142K}, J0125+0759 \citep{2020ApJ...898..142K}, that are observable in the given Keck/LRIS nights. These are bright EMPGs whose faint emission lines, especially [Fe {\sc iii}]$\lambda$4658, can be potentially detected.
We can detect the other important auroral lines that are necessary to discuss the abundance ratios (e.g. {\sc [O iii]}$\lambda$4363 and {\sc [S iii]}$\lambda$6312; see Section \ref{sec:abundance}).

\subsection{Keck/LRIS Spectroscopy} \label{sec:obs_Spec}

We conducted spectroscopic observations for the EMPGs in 2021 November 7 and 8 with Keck/LRIS (PI: K.Nakajima).
LRIS has the blue and red channels that cover the wavelength ranges of $\lambda \sim 3000-5500$ and $6000-9000$ \AA\ with the spectral resolutions of $\sim 4$ and 5 \AA\ in FWHM, respectively.
We used the 600 lines mm$^{-1}$ grism blazed at 4000 \AA\ on the blue channel and the 600 lines $\mathrm{mm}^{-1}$ grating blazed at 7500 \AA\ on the red channel. 
The slit widths were $0^{\prime \prime}.7$ for all targets.
We also observed spectrophotometric standards Feige 34 and Feige 100 for flux calibration.
The sky was clear during the observations with seeing sizes of $0''.8-1''.0$.

\begin{table*}[t]
    \begin{center}
    \tabletypesize{\scriptsize}
    \tablewidth{0pt} 
    \caption{Obseravation Targets}
    \begin{tabular}{cccc}\hline \hline
        ID & R.A. & Dec. & Exposure\\
           & hh:mm:ss & dd:mm:ss  & s\\
     (1) & (2) & (3) & (4) \\ \hline
    SBS-0335-052E & 03:37:44.0 & -05:02:40.0 & 1800 \\
    J2314+0154 & 23:14:37.6 & +01:54:14.3 & 1200 \\
    J0125+0759 & 01:25:34.2 & +07:59:24.7 & 1200 \\ \hline
    \end{tabular}
    
\tablecomments{(1) ID. (2) Right ascension in J2000. (3) Declination in J2000.(4) Slit width. (5) Exposure time.}
\label{table_obs}
\end{center}
\end{table*}

\subsection{Reduction} \label{sec:obs_reduction}
We reduce the LRIS data using the IRAF package in a normal manner, performing bias subtraction, flat-fielding, cosmic-ray cleaning, sky subtraction, wavelength calibration, one-dimensional (1D) spectrum extraction, flux calibration, atmospheric-absorption correction, and Galactic-reddening correction.
A 1D spectrum is derived from an aperture centered on the compact component of our galaxies. 
The 1D spectra are corrected for Galactic extinction according to the spatial position of each object on the \cite{2011ApJ...737..103S}'s dust map and based on the extinction curve of \cite{1989ApJ...345..245C}, as commonly performed for extragalactic objects.
Flux calibration is obtained from the spectrophotometric standard stars observed on the same night under a similar seeing condition with the same slit width ($0''.7$), and reduced in the same manner as done for the science targets.
 The wavelengths are calibrated with the HgNeArCdZnKrXe lamp.
 Atmospheric absorption is corrected under the assumption of the extinction curve at Mauna Kea Observatories.
In addition, the flux calibration using the spectrophotometric standard stars eliminates local absorption features due to the earth's atmosphere.

\begin{figure*}[ht]
\centering
\includegraphics[width=16cm]{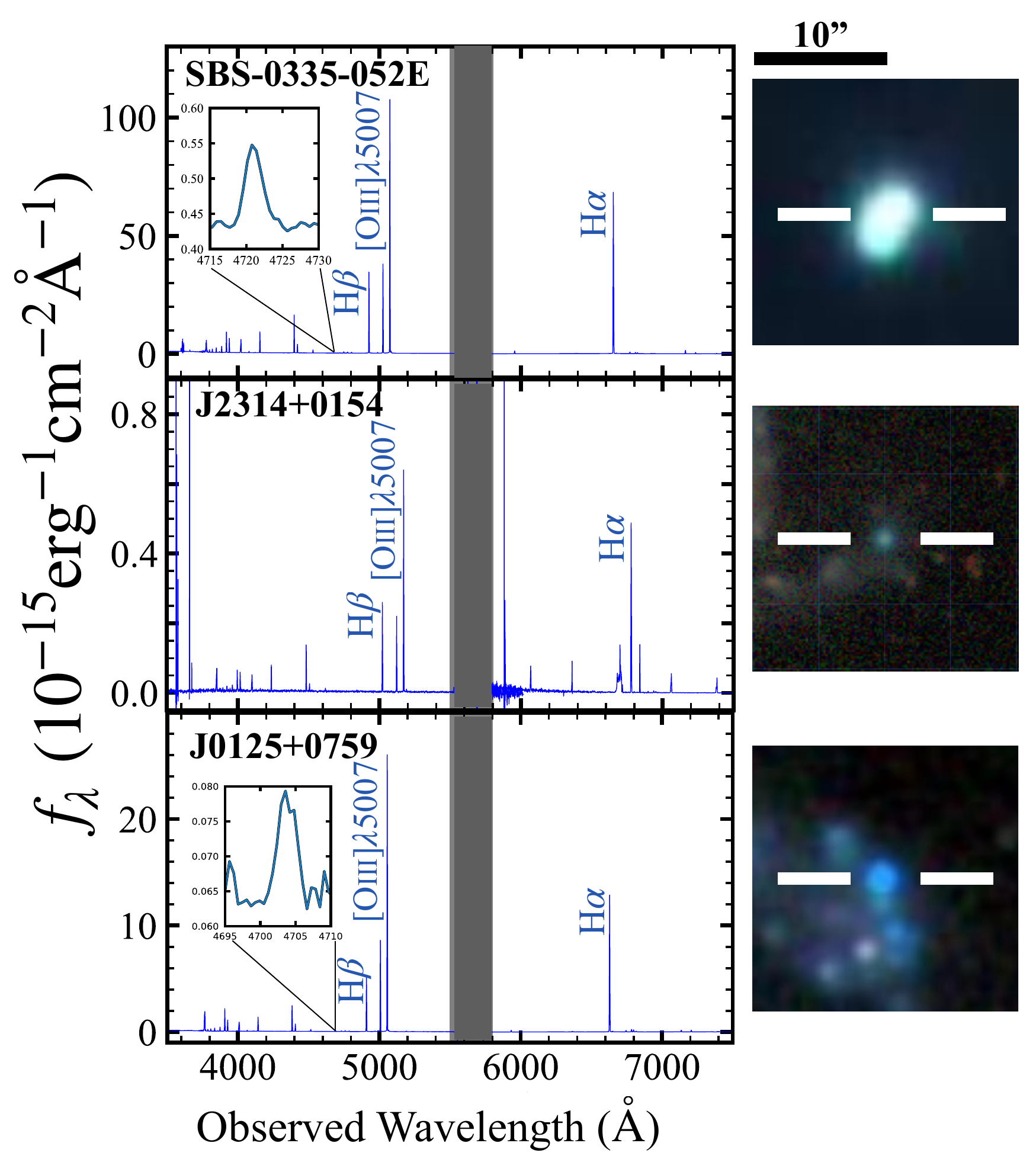}
\caption{LRIS spectrum of our targets. The top, middle, and bottom panels present SBS-0335-052E, J2314+0154, and J0125+0759, respectively.
The gray shade region indicates the gap between the LRIS blue and red channels.
The top-right, middle-right, and bottom-right panels show $20''\times 20''$ cutout $gri-$composite images of SBS-0335-052E, J2314+0154, and J0125+0759, from Pan-STARRS1, HSC, and SDSS, respectively.
The inset panels at the top-left corner of each panel represent the enlarged view of the spectrum around [Fe{\sc iii}] $\lambda$4658. In J2314+0154, the emission line of [Fe{\sc iii}] $\lambda$4658 is not detected.
\label{fig:spectra}}
\end{figure*}

Figure \ref{fig:spectra} represents the reduced spectra of our targets.
In the middle panels, the spectrum of J2314+0154 exhibits unusual emission lines near H$\mathrm{\alpha}$ due to the incomplete removal of certain sky emission lines. Nevertheless, these sky emission lines do not coincide with the emission lines of the object.
 
\subsection{Emission Line Measurements}

We measure the emission line flux by fitting a Gaussian profile plus a constant continuum using the {\tt scipy.optimize} package \citep{2020NatMe..17..261V}.
We apply the $\chi^2$ minimization approach considering the error spectra.
The error spectra are extracted by taking into account the readout noise and photon noise from both sky and object counts.
In the Gaussian profile fitting, there are four free parameters: the amplitude, line width, line central wavelength, and the continuum.

To perform reddening-corrections for the observed fluxes, we estimate dust extinction from the Balmer decrements under the assumptions of the case B \citep{1971MNRAS.153..471B} recombination and the dust attenuation curve given by \cite{2000ApJ...533..682C}.
We estimate intrinsic Balmer decrement values using PyNeb \citep{2015A&A...573A..42L}.
We summarize the atomic data used in the Pyneb calculation in Table \ref{table_atomic}.
%
Because the Balmer decrement values depend on electron temperature $T_\mathrm{e}$ and electron density $n_\mathrm{e}$, 
we iteratively derive $E(B-V)$ values that consistently explain $T_\mathrm{e}$ and $n_\mathrm{e}$ (See Section \ref{sec:abundance} for the procedures of $T_\mathrm{e}$ and $n_\mathrm{e}$ calculations).
We utilize 6 Balmer line ratios, H$\mathrm{\beta}$/H$\mathrm{\alpha}$, H$\mathrm{\gamma}$/H$\mathrm{\alpha}$, H$\mathrm{\delta}$/H$\mathrm{\alpha}$, H$\mathrm{\gamma}$/H$\mathrm{\beta}$, H$\mathrm{\delta}$/H$\mathrm{\beta}$, and H$\mathrm{\gamma}$/H$\mathrm{\delta}$ to estimate the emissivity of the Balmer lines. 
We compare the Balmer line ratios of the observational measurements with those of theoretical predictions obtained with PyNeb to determine $E(B-V)$.
With the $E(B-V)$ values thus obtained,
we determine the best-estimate $E(B-V)$ by the $\chi^2$ minimization for each  Balmer decrement value.
We also estimate $\pm 68\ \%$ confidence intervals of $E(B - V )$ based on $\chi^2$.
With the $E(B - V )$ values and the attenuation curve \citep{2000ApJ...533..682C},
we correct all of the observed emission line fluxes for dust extinction
that are summarized in Table \ref{table:flux_list}.
Table \ref{table_sample} summarizes the fundamental properties of our targets such as the redshift, and $E(B-V)$.

\begin{table*}
    \begin{center}
    \tabletypesize{\scriptsize}
    \tablewidth{0pt} 
    \caption{Atomic Data}
    \begin{tabular}{cccc} \hline \hline
    Ion & Emission Process &  Transition Probability & Collision Strength \\
    (1) & (2) & (3) & (4)  \\ \hline
    $\mathrm{H^0}$ & Re & \cite{atom_H0} & -  \\
    $\mathrm{O^{+}}$ & CE & \cite{FFT04} & \cite{atom_O2_coll} \\ 
    $\mathrm{O^{2+}}$ & CE  & \cite{FFT04} & \cite{atom_O3_coll} \\
    $\mathrm{Fe^{2+}}$ & CE & \cite{atom_fe_Q},\cite{atom_fe_J} & \cite{atom_fe_Z} \\
    $\mathrm{Ar^{2+}}$ & CE & \cite{atom_ar} & \cite{atom_ar} \\ 
    $\mathrm{Ar^{3+}}$ & CE  & \cite{atom_RGJ19} & \cite{atom_Ar4_coll} \\
    $\mathrm{S^+}$ & CE & \cite{atom_RGJ19} & \cite{atom_S2_coll}  \\
    $\mathrm{S^{2+}}$ & CE & \cite{atom_S3} & \cite{atom_S3_coll} \\ 
    $\mathrm{Ne^{2+}}$ & CE  & \cite{FFT04} & \cite{atom_Ne_coll} \\
    $\mathrm{N^{2+}}$ & CE & \cite{FFT04} & \cite{atom_N_coll}  \\ \hline
 \end{tabular}
\tablecomments{(1) Ion. (2) Emission process. Re and CE represent recombination and collisional excitations, respectively. (3) Reference of the transition probability. (4) Reference of the collision strength.}
\label{table_atomic}
\end{center}
\end{table*}

\begin{splitdeluxetable*}{lcccccccccBccccccccc}
\tablecaption{Dust-corrected Fluxes of Our Targets
\label{table:flux_list}}
\renewcommand{\arraystretch}{1.1}
\tabletypesize{\scriptsize}
\tablehead{
\colhead{ID} &\colhead{[O{\sc ii}]$\lambda\lambda$ 3727,3729} &
\colhead{[Ne{\sc iii}] $\lambda$3869} &\colhead{$\mathrm{H\delta}$} &
\colhead{$\mathrm{H\gamma}$} &
\colhead{[O{\sc iii}] $\lambda$4363} & \colhead{[Fe{\sc iii}] $\lambda$4658}  &
\colhead{[Ar{\sc iv}] $\lambda$4711 + He~{\sc i}}&
\colhead{[Ar{\sc iv}] $\lambda$4740} & $\mathrm{H\beta}$ &
\colhead{[O{\sc iii}] $\lambda$4959} &\colhead{[O{\sc iii}] $\lambda$5007} &
\colhead{[S{\sc iii}] $\lambda$6312} &
\colhead{$\mathrm{H\alpha}$} &\colhead{[N{\sc ii}] $\lambda$6583} &
\colhead{[S{\sc ii}] $\lambda$6716} &
\colhead{[S{\sc ii}] $\lambda$6731} &
\colhead{[Ar{\sc iii}] $\lambda$7136} & \colhead{$F(\mathrm{H\beta})$}
\\ 
& 
& & 
& & 
& &
& & 
& & 
& & 
& & 
& & 
 &$10^{-16} \mathrm{erg} \mathrm{s}^{-1} \mathrm{~cm}^{-2}$
}
\colnumbers
\startdata
 SBS-0335-052E & $22.02 \pm 0.02$  & $24.88 \pm 0.03$ & $22.53 \pm 0.03$ & $47.89 \pm 0.07$ & $10.81 \pm 0.04$ & $0.32 \pm 0.03 $ & $1.70 \pm 0.03$ & $0.91 \pm0.03 $ & $100.00 \pm 0.17$ & $103.52 \pm 0.19$ & $301.26 \pm 0.32$ 
 & $0.34 \pm 0.06$ &  $220.16 \pm 0.75$ & $0.69 \pm 0.1$ & $1.47 \pm0.10$ & $1.17 \pm 0.10$ & $1.78 \pm 0.10$ & $972.1 \pm 16.5$ \\ 
J2314+0154 & $51.09 \pm 0.65$ & $28.14 \pm 0.51$ & $31.61 \pm 0.54$ & $54.70 \pm 0.81$ & $8.02 \pm 0.51$ & $\dots$ &$\dots$ &$\dots$ & $100.00 \pm 1.66$ & $80.66 \pm 1.63$ & $233.40 \pm 2.79$ & $\dots$ & $256.11 \pm 26.15$ & $ \dots$ & $3.02 \pm 6.73$& $2.22 \pm 8.28$ &$\dots$ & $7.279 \pm 1.206$ \\ 
 J0125+0759 & $62.09 \pm 0.05$ & $38.29 \pm 0.06$ & $27.07 \pm 0.10$ & $50.45 \pm 0.14$ & $14.22 \pm 0.15$ & $0.28 \pm 0.20$ & $1.93 \pm 0.23$ & $1.04 \pm 0.21$ & $100.00 \pm 0.25$ & $163.27 \pm 0.24$ & $484.45 \pm 0.24$ & $0.71 \pm 0.10$ & $276.45 \pm 0.14$ & $\dots$ & $4.60 \pm 0.12$ &$4.07 \pm 0.13$ & $3.43 \pm 0.23$ 
 & $164.7 \pm  0.40 $\\
\enddata
\tablecomments{(1) ID. (2)-(18) Dust corrected flux normalized by $\mathrm{H\beta}$ as 100. (19)Aperture-corrected flux of $\mathrm{H\beta}$.
}
\end{splitdeluxetable*}

\begin{table*}
    \begin{center}
    \tabletypesize{\scriptsize}
    \tablewidth{0pt} 
    \caption{Fundamental Properties of Our Targets}
    \begin{tabular}{cccccc} \hline \hline
    ID &  Redshift &  $E(B-V)$ &  $T_e$({\sc [Oiii]}) &  $n_e$&    12+$\log (\mathrm{O/H})$ \\
    &  & (mag) &  ($10^4$~K) &  ($\mathrm{cm^{-3}}$) & \\
    (1) & (2) & (3) & (4) & (5) & (6) \\ \hline
 SBS-0335-052E & 0.0135 & $0.031 \pm 0.027$ & $2.076 \pm 0.005$ & $260.72^{+97.33}_{-81.05}$ & $7.23 \pm 0.01^\dagger$ \\
 J2314+0154 & 0.0327 & $0.123 \pm 0.019$ & $2.021^{+0.388}_{-0.336}$ & $83.47^{+301}_{-83}$ & $7.19 \pm 0.01$\\ 
 J0125+0759 & 0.010  & $0.067 \pm 0.035$  & $1.844 \pm 0.004$ & $504.12^{+52.85}_{-49.90}$ & $7.56 \pm 0.01$\\ \hline
 \end{tabular}
\tablecomments{(1) ID. (2) Redshift. (3) $E(B-V)$. (4) Electron temperature of $T_\mathrm{e}${(\sc{[Oiii]})} . (5) Electron density $n_\mathrm{e}$ ({\sc{[Sii]}})  (6) The gas phase metallicity measured as 12+$\log$(O/H). \\
$\dagger$ 
The uncertainty, 0.01, is the upper limit.
} 
\label{table_sample}
\end{center}
\end{table*}

\section{Sample} \label{sec:sample}
\subsection{Developing our Sample}\label{subsec:sample}

We investigate the Keck/LRIS spectra of our three EMPGs (Section \ref{sec:obs_Spec}).
Although many lines of Ar, S, Ne, N, and O emission are identified in the spectra,
only two out of three EMPGs, SBS-0335-052E, and J0125+0759, have a significant detection of [FeIII] emission.
We hereafter use SBS-0335-052E and J0125+0759 in our analysis.
We also use 11 EMPGs with the detections of the Fe, Ar, S, Ne, N, and O emission lines, taken from the literature \citep{2022ApJ...925..111I,2021ApJ...913...22K,2018MNRAS.478.4851I,1999ApJ...511..639I,2021ApJ...922..170B}.
We define our sample that consists of a total of the 13 EMPGs from our observations and the literature, and summarize our sample in Table \ref{table_abundance}.

\begin{deluxetable*}{ccccccc}
\tabletypesize{\scriptsize}
\tablewidth{0pt}
\tablecaption{Our Sample}
\tablehead{\colhead{Name} & \colhead{[Fe/O]} & \colhead{[Ar/O]} &\colhead{[S/O]} &\colhead{[Ne/O]} &\colhead{[N/O]} & \colhead{Reference}}
\colnumbers
 \startdata 
 SBS-0335-052E& $-0.24^{+0.11}_{-0.24}$ & {$-0.06 \pm 0.01$ }& $-0.18^{+0.05}_{-0.018}$ & $-0.27 \pm 0.00$  & $-0.63^{+0.05}_{-0.05}$  & This paper \\
 J0125+0759& $-0.24^{+0.12}_{-0.24}$ & {$-0.08 ^{+0.02}_{-0.06}$} & $-0.34^{+0.05}_{-0.05}$ & $-0.01\pm 0.00$ & --  & This paper \\
 J0159-0622 & $-0.51^{+0.17}_{-0.28}$ & $-0.01\pm 0.02$ & $-0.016 \pm 0.06$ & $-0.12^{+0.01}_{-0.02}$ & $-0.61 \pm 0.03$ &\cite{2022ApJ...925..111I} \\ 
 J1608+4337 & $-0.43^{+0.15}_{-0.22}$ & $-0.04^{+0.02}_{-0.03}$ & $-0.07 \pm 0.02$ & $0.107 \pm 0.02$ & $-0.61^{+0.02}_{-0.03}$ &\cite{2022ApJ...925..111I} \\ 
 J1631+4426 & $-0.02^{+0.17}_{-0.31}$ & -  & $-0.205^{+0.05}_{-0.04}$ & $0.119^{+0.022}_{-0.019}$ & $< -0.85$ &\cite{2021ApJ...913...22K}\\
 J2115-1734 & $-0.41^{+0.03}_{-0.02}$ & $0.04^{+0.01}_{-0.06}$ & $-0.079 \pm 0.06$ & $0.003^{+0.004}_{-0.005}$ & $-0.658^{+0.009}_{-0.011}$&\cite{2021ApJ...913...22K}\\ 
 J0811+4730 & $0.17 \pm 0.092$ & $-0.245 \pm 0.104$ & $-0.117 \pm 0.056$ & $-0.0048 \pm 0.031$ & $-0.675 \pm 0.044$  &\cite{2018MNRAS.478.4851I}\\
 I Zw 18 & $-0.38 \pm 0.12$ & $0.01 \pm 0.03$ & $-0.08 \pm 0.05$ & $-0.02 \pm 0.03$ & $-0.74 \pm 0.03$  &\cite{1999ApJ...511..639I}\\
 0940+544N & $-0.34 \pm 0.04$ & $0.09 \pm 0.03$ & $-0.04 \pm 0.03$ & $0.04 \pm 0.03$ & $-0.75 \pm 0.03$  &\cite{1999ApJ...511..639I}\\
 1159+545 & $-0.14 \pm 0.05$ & $0.18 \pm 0.05$ & $0.06 \pm 0.04$ & $0.03 \pm 0.04$ & $-0.72 \pm 0.04$  &\cite{1999ApJ...511..639I}\\
 1211+540 & $-0.47 \pm 0.02$ & $-0.08 \pm 0.02$ & $0.09 \pm 0.02$ & $0.01 \pm 0.02$ & $-0.73 \pm 0.02$  &\cite{1999ApJ...511..639I}\\
 J104457 & $-0.26 \pm 0.12$ & $-0.02 \pm 0.07$ & $-0.23 \pm 0.09$ & $-0.03 \pm 0.03$ & $-0.57 \pm 0.10$  &\cite{2021ApJ...922..170B}\\
 J141851 & $-0.45 \pm 0.08$ & $0.11 \pm 0.05$ & $-0.37 \pm 0.07$ & $0.01 \pm 0.03$ & $-0.80 \pm 0.07$  &\cite{2021ApJ...922..170B}\\
 \enddata
\tablecomments{(1) ID. (2)-(6) Abundance ratios normalized by solar abundances \citep{Asplaud} (7) Reference. }
\label{table_abundance}
\end{deluxetable*}

\subsection{Element Abundance Ratios}\label{sec:abundance}
We derive the oxygen abundances with the direct method.
We estimate the electron temperature $T_\mathrm{e}$ of doubly-ionized oxygen $T_\mathrm{e}(\mathrm{OIII})$
from ratios of two collisional excitation line fluxes that depend on $T_\mathrm{e}(\mathrm{OIII})$.
We use the PyNeb package {\tt getCrossTemDen} to derive simultaneously $n_\mathrm{e}$ and $T_\mathrm{e}(\mathrm{OIII})$ from emission-line ratios of {\sc [Sii]}$\lambda$6731/ {\sc [Sii]}$\lambda$6716 and {\sc [Oiii]}$\lambda$4363/ {\sc [Oiii]}$\lambda\lambda$ 4959,5007, respectively. 
Table \ref{table_sample} summarizes our estimates of $n_\mathrm{e}$ and $T_\mathrm{e}$ for all of our galaxies.

We use the PyNeb package {\tt getIonAbundance} to obtain ion abundance ratios.
Ion abundance ratios of O$^{2+}$/H$^+$ and O$^{+}$/H$^+$ are derived from emission-line ratios of {\sc [Oiii]}$\lambda\lambda$4959,5007/H$\beta$ and {\sc [Oii]}$\lambda\lambda$3727,3729/H$\beta$ with $T_\mathrm{e}${\sc (Oiii)} and $T_\mathrm{e}${\sc (Oii)}, respectively, 
where $T_\mathrm{e}${\sc (Oii)} is the electron temperature of singly-ionized oxygen that is
estimated with the empirical relation of
\begin{equation}
T_{\mathrm{e}}(\mathrm{O}\text{II})=0.7 \times T_{\mathrm{e}}(\mathrm{O}\text {III})+3000
\end{equation}
\citep{1992AJ....103.1330G}.
By adding O$^{2+}$/H$^+$ and O$^{+}$/H$^+$, we obtain the total oxygen abundance O/H.
We represent the metallicity as 12+$\log$(O/H) of the total oxygen abundance.

We derive ion abundance ratios of Fe$^{2+}$/H$^+$, Ar$^{2+}$/H$^+$, 
S$^{2+}$/H$^+$, S$^{+}$/H$^+$, N$^+$/H$^+$, and Ne$^{2+}$/H$^+$ that are estimated from the fluxes of  
[Fe{\sc iii}]$\lambda$4658, [Ar{\sc iii}]$\lambda$7136, 
{\sc [Siii]}$\lambda$6312, {\sc [Sii]}$\lambda\lambda$6716,6731, {\sc [Nii]}$\lambda\lambda$6548,6583, and,  [Ne{\sc iii}]$\lambda$3869, respectively, with the electron temperatures.
Here we use $T_\mathrm{e}${\sc (Oii)} for Fe$^{2+}$, S$^{+}$, and N$^{+}$, because ionization potential energies of them are 10.4--16.2 eV close to 13.6 eV which is the ionization potential energy of O$^{+}$\citep{2021ApJ...922..170B}.
Similarly, we apply $T_\mathrm{e}${\sc (O iii)} for Ne$^{2+}$ 
since the ionization energy of Ne$^{2+}$ 
is 
41.0 eV comparable with the one of O$^{++}$.
Because the ionization energies of Ar$^{2+}$ and 
S$^{2+}$ range in 23.3--27.6 eV significantly lower than the one of O$^{++}$, we do not use $T_\mathrm{e}${\sc (O iii)} for Ar$^{2+}$ and S$^{2+}$. 
Instead, we use the electron temperature of S$^{2+}$, $T_\mathrm{e}${\sc (S iii)}, that is
estimated by the empirical relation of
\begin{equation}
T_{\mathrm{e}}(\mathrm{S}\text{III})=0.83 \times T_{\mathrm{e}}(\mathrm{O}\text {III})+1700
\end{equation}
\citep{1992AJ....103.1330G}.
We use the ionization correction factors (ICFs) derived by \cite{2006A&A...448..955I} to calculate total gas-phase element abundances from the ion abundances:
\begin{equation}
\frac{\mathrm{Fe}}{\mathrm{H}}=\frac{\mathrm{Fe}^{2+}}{\mathrm{H}^{+}} \times \operatorname{ICF}\left(\mathrm{Fe}^{2+}\right),
\label{eq:Fe_H}
\end{equation}
\begin{equation}
\frac{\mathrm{Ar}}{\mathrm{H}}=\frac{\mathrm{Ar}^{2+}}{\mathrm{H}^{+}} \times \operatorname{ICF}\left( \mathrm{Ar}^{2+}\right),
\end{equation}
\begin{equation}
\frac{\mathrm{S}}{\mathrm{H}}=\frac{\mathrm{S}^{2+}+\mathrm{S}^{+}}{\mathrm{H}^{+}} \times \operatorname{ICF}\left(\mathrm{S}^{2+} + \mathrm{S}^{+}\right),
\end{equation}
\begin{equation}
\frac{\mathrm{N}}{\mathrm{H}}=\frac{\mathrm{N}^{+}}{\mathrm{H}^{+}} \times \operatorname{ICF}\left(\mathrm{N}^{+}\right),
\end{equation}
\begin{equation}
\frac{\mathrm{Ne}}{\mathrm{H}}=\frac{\mathrm{Ne}^{2+}}{\mathrm{H}^{+}} \times \operatorname{ICF}\left(\mathrm{Ne}^{2+}\right).
\label{eq:Ne_H}
\end{equation}
The ICFs depend on the ionization degree of gas measured with O$^+$ and O$^{2+}$:
\begin{equation}\label{eq:ICF_Fe}
\begin{aligned}
\operatorname{ICF}\left(\mathrm{Fe}^{2+}\right) &=0.036 v-0.146+1.386 / v, \text { low } Z, \\
&=0.301 v-0.259+1.367 / v, \text { intermed. } Z, \\
&=-1.377 v+1.606+1.045 / v, \text { high } Z,
\end{aligned}
\end{equation}
\begin{equation}
\begin{aligned}
\operatorname{ICF}\left(\mathrm{Ar}^{2+}\right) &=0.278 v+0.836+0.051 / v, \text { low } Z, \\
&=0.285 v+0.833+0.051 / v, \text { intermed. } Z, \\
&=0.517 v+0.763+0.042 / v, \text { high } Z,
\end{aligned}
\end{equation}
\begin{equation}
\begin{aligned}
\operatorname{ICF}\left(\mathrm{~S}^{+}+\mathrm{S}^{2+}\right) &=0.121 v+0.511+0.161 / v, \text { low } Z, \\
&=0.155 v+0.849+0.062 / v, \text { intermed. } Z, \\
&=0.178 v+0.610+0.153 / v, \text { high } Z,
\end{aligned}
\end{equation}
\begin{equation}
\begin{aligned}
\operatorname{ICF}\left(\mathrm{~N}^{+}\right) &=-0.825 v+0.718+0.853 / v, \text { low } Z, \\
&=-0.809 v+0.712+0.852 / v, \text { intermed. } Z, \\
&=-1.476 v+1.752+0.688 / v, \text { high } Z,
\end{aligned}
\end{equation}
\begin{equation}\label{eq:ICF_Ne}
\begin{aligned}
\operatorname{ICF}\left(\mathrm{Ne}^{2+}\right) &=-0.385 w+1.365+0.022 / w, \text { low } Z, \\
&=-0.405 w+1.382+0.021 / w, \text { intermed. } Z, \\
&=-0.591 w+0.927+0.546 / w, \text { high } Z,
\end{aligned}
\end{equation}
where $v$ and $w$ are $\mathrm{O}^{+}/(\mathrm{O}^{2+}+\mathrm{O}^{+})$ and $ \mathrm{O}^{2+}/(\mathrm{O}^{2+}+\mathrm{O}^{+})$, respectively \citep{2006A&A...448..955I}.
In the equations, `low Z', `intermed. Z', and `high Z' indicate $12+\log (\mathrm{O} / \mathrm{H}) \leqslant 7.2$, $7.2<12+\log (\mathrm{O} / \mathrm{H})<8.2$, and $8.2 \leqslant12+\log (\mathrm{O} / \mathrm{H})$, respectively.
We apply 'low Z' ('intermed Z') to J2314+0154 (SBS-033-052E and J0125+0759).
Dividing the values of Eqs. (\ref{eq:Fe_H})- (\ref{eq:Ne_H}) by O/H, we obtain Fe/O, Ar/O, S/O, N/O, and Ne/O.
We use the Ar/O values derived with [Ar{\sc iii}]$\lambda$7136 emission because [Ar{\sc iv}]$\lambda$4711 emission lines are contaminated by He~{\sc i} emission lines. 
Although our best estimate Ar/O values are obtained with [Ar{\sc iii}]$\lambda$7136, we examine these Ar/O values with those estimated from the [Ar{\sc iv}]$\lambda$4711, [Ar{\sc iv}]$\lambda$4740, and [Ar{\sc iii}]$\lambda$7136 emission lines albeit with the He~{\sc i} emission contamination.
We confirm that the best estimate Ar/O values are consistent with the Ar/O upper limits that are the Ar/O values derived with the He~{\sc i} emission contamination.

Here we estimate the errors of element abundance ratios.
Conducting Monte-Carlo simulations, 
we generate 1000 mock flux values 
consisting of the observed flux and a flux
randomly produced on the basis of the normal distribution 
whose dispersion corresponds to the $1\sigma$ flux error.
We obtain 1000 mock element abundance ratios from the mock flux values, 
and define the $1\sigma$ error of the element abundance ratio
as the 68\% confidence interval in the distribution of the 1000 mock element abundance ratios.

Table \ref{table_abundance} summarizes the gas-phase element abundance ratios for all of our galaxies.
We use the abundance ratios taken from the literature \citep{2022ApJ...925..111I,2021ApJ...913...22K,2018MNRAS.478.4851I,1999ApJ...511..639I,2021ApJ...922..170B} that derive the abundance ratios in the same manner as our methods.
The abundance ratios taken from \cite{1999ApJ...511..639I} and \cite{2021ApJ...922..170B} are derived from the ICFs different from the equations (\ref{eq:ICF_Fe}) - (\ref{eq:ICF_Ne}).
We calculate the abundance ratios for the galaxies in \cite{1999ApJ...511..639I} and \cite{2021ApJ...922..170B} using the fluxes reported by \cite{1999ApJ...511..639I} and \cite{2021ApJ...922..170B}  and the ICFs of the equations (\ref{eq:ICF_Fe}) - (\ref{eq:ICF_Ne}).
Because these abundance ratios are consistent with the values derived by \cite{1999ApJ...511..639I} and \cite{2021ApJ...922..170B} within the errors, we use the abundance ratios reported by \cite{1999ApJ...511..639I} and \cite{2021ApJ...922..170B}.
Because there are no S/O values of the four galaxies derived in \cite{2021ApJ...913...22K} and \cite{2022ApJ...925..111I}, 
we calculate the S/O values from the fluxes reported by \cite{2021ApJ...913...22K} and \cite{2022ApJ...925..111I}.
We assume that the EMPGs are not affected by dust depletion since the dust in the EMPGs is poor due to low metallicities in EMPGs. 
Therefore, we have not corrected dust depletion in the gas-phase element abundance ratios.
We discuss the effect on the abundance ratios quantitatively later.

\section{Chemical Evolution Models} \label{sec:model}
\subsection{Yield Models}\label{sec:yield}

We calculate yields with the models of \citet{2007ApJ...660..516T}. We use these yields 
in the galactic chemical evolution models (Section \ref{sec:model_develop}).
We investigate the yields for EMPGs that eject rich iron without increasing sulfur and argon.

\subsubsection{CCSN \& HN Yields}

We calculate CCSN and HN yields with the explosive nucleosynthesis code \citep{2007ApJ...660..516T} in order to study the origin of enriched Fe in the EMPGs. We obtain the yields by using calculation code and explosive nucleosynthesis from \cite{2007ApJ...660..516T}.
 We calculate the yields with different parameters.
 We use progenitor initial masses (13, 15, 18, 20, 25, 30, and 40 $M_\odot$) and explosion energies (CCSNe with $E_{51} = E/10^{51} \mathrm{erg} = 1$, and HNe with $E_{51} \geq 10$).
 The explosion energies are determined by the relationship between the main-sequence mass and the explosion energy which is obtained from observations and supernova models \citep{2007ApJ...660..516T}.
 Since we compare with the EMPGs observations, the metallicities of yields are set to $Z=0$ and $0.004(=0.288~Z_\odot)$.

We apply the mixing \& fallback model proposed by \citet{2003Natur.422..871U,2002ApJ...565..385U}.
This model is introduced to reproduce the abundance ratios of metal-poor stars.
In this model, we assume that inner materials in the mixing region are mixed during supernovae by some mixing process (e.g., Rayleigh- Taylor instabilities or aspherical explosions).
Then some fraction of the material in the mixing region is ejected into interstellar space and the rest undergoes fallback to the center remnant due to gravity.
This model can modify the abundance ratios of the material released by supernova explosions through the mixing and fallback processes. 
We investigate whether the yields with the mixing \& fallback model may reproduce the characteristic abundance ratios of EMPGs.

The mixing region and the amount of ejecta and fallback are described by the following parameters: $M_\mathrm{cut}$, $M_\mathrm{mix}$, and $f_\mathrm{ej}$.
The initial mass cut $M_\mathrm{cut}$ represents the inner boundary of the mixing region. 
$M_\mathrm{mix}$ is the outer boundary of the mixing region.
All material above $M_\mathrm{mix}$ is ejected.
$M_\mathrm{cut}$ and $M_\mathrm{mix}$ indicate the enclosed masses from the center of a star.
It is likely to be difficult to eject the iron core
due to energy absorption by Fe photodisintegration.
$f_\mathrm{ej}$ is the ejection fraction.
A fraction $f_\mathrm{ej}$ of the material in the mixing region (between $M_\mathrm{mix}$ and $M_\mathrm{cut}$) is ejected into interstellar space.
We adopt $f_\mathrm{ej} = 0.12$, following \cite{2007ApJ...660..516T}.
We change the $f_\mathrm{ej}$ from 0.12 to 0.05 and 0.5 and calculate the yields to examine the impact of $f_\mathrm{ej}$ variation on the models. We find that the variations in $f_\mathrm{ej}$ do not change our conclusion.

$M_\mathrm{mix}$ is expressed by
\begin{equation}
M_{\mathrm{mix}}=M_{\mathrm{cut}}+x\left(M_{\mathrm{CO}}-M_{\mathrm{cut}}\right),
\end{equation}
where $M_\mathrm{CO}$ is  CO core mass and $x$ is the mixing region factor \citep{2018ApJ...857...46I}.
Here, we change the mixing region by varying $x$.
If $x=0$, the mixing region $M_\mathrm{mix}$ equals $M_\mathrm{cut}$ and all material above the Fe core is released.
We set $x=0$, 0.1, 0.2, 0.5, and 1.0 to vary the mixing region, because the target elements of this study are mainly located in the CO core.
\cite{N13papaer} extends the mixing region outside of the Si-burning layer, where an $x$ value simply depends on mass. In our study, we extensively investigate the mixing and fallback effects with multiple $x$ values for a given mass, because the yields of the literature do not explain the abundance ratios of EMPGs.
Table \ref{table_yield} summarizes the parameters of yields.

Then, we develop the models about CCSNe and HNe using the calculated yields (see section \ref{sec:model_develop}).

\begin{deluxetable*}{ccccc}
\tablecaption{Parameters of SN yields \label{tab:deluxesplit}}
\tabletypesize{\scriptsize}
\tablehead{
\colhead{{$M/M_\odot$}}
&\colhead{$M_\mathrm{CO}~/M_\odot$}
&\colhead{$M_\mathrm{Fe}~/M_\odot$}
&\colhead{$E_\mathrm{CCSN}~/10^{51}~\mathrm{erg}$}
&\colhead{$E_\mathrm{HN}~/10^{51}~\mathrm{erg}$}
}
\colnumbers
\startdata
\multicolumn{5}{c}{$Z=0$}\\
\hline
13 & 2.39 & 1.47 & 1 & 10 \\
15 & 3.02 & 1.41 & 1 & 10 \\
18 & 4.15 & 1.54 & 1 & 10 \\
20 & 5.28 & 1.53 & 1 & 10 \\
25 & 6.29 & 1.69 & 1 & 10 \\
30 & 8.75 & 1.85 & 1 & 20 \\
40 & 13.89 & 2.42 & 1 & 30 \\
\hline
\multicolumn{5}{c}{$Z=0.004$}\\
\hline
13 & 2.37 & 1.52 & 1 & 10 \\
15 & 2.24 & 1.40 & 1 & 10 \\
18 & 2.59 & 1.52 & 1 & 10 \\
20 & 3.43 & 1.58 & 1 & 10 \\
25 & 5.26 & 1.58 & 1 & 10 \\
30 & 9.38 & 2.10 & 1 & 20 \\
40 & 14.46 & 2.46 & 1 & 30 \\
\enddata
\tablecomments{(1) Progenitor mass. (2) CO core mass. (3) Fe core mass. (4) Explosion energy of CCSN. (5) Explosion energy of HN. }
\label{table_yield}
\end{deluxetable*}

\subsection{Development of the Galactic Chemical Evolution Models}\label{sec:model_develop}

To understand the origin of the high Fe/O, we compare the galactic chemical evolution models with the observations.
\cite{2022ApJ...925..111I} construct Fe/O evolution models based on \cite{2018ApJ...852..101S}.
We develop models about Ar/O, S/O, N/O, and Ne/O in the same way.

These models are one-box chemical evolution models.
We assume instantaneous star formation.
We create stars based on an IMF of \cite{2001MNRAS.322..231K} which is expressed by the broken power-law function $\Phi(M) \propto M^{-\alpha}$ with 
\begin{equation}
\alpha=\left\{\begin{array}{l}
0.3 \text { for } M / M_{\odot}<0.08, \\
1.3 \text { for } 0.08 \leqslant M / M_{\odot}<0.5, \\
2.3 \text { for } 0.5 \leqslant M / M_{\odot}.
\end{array}\right.\\
\end{equation}
Table \ref{table_literaturemodel} presents the mass ranges for each model. We determine the mass ranges of our models, following \cite{2018ApJ...852..101S}.

\begin{deluxetable*}{ccccc}[htp]
\tabletypesize{\scriptsize}
\tablewidth{0pt} 
\tablecaption{Models of the literature yields \label{tab:deluxesplit}}
\tablehead{\colhead{Model} & \colhead{Supernova}& \colhead{Progenitor Star} & \colhead{ Yields } & \colhead{Reference}}
\colnumbers
 \startdata 
 { } & { } & { $9 ~ M_\odot \leq  M < 13~M_\odot$} & {extrapolation} & { } \\
 {   }& {    } & {$13 ~ M_\odot \leq  M \leq 40~M_\odot$} & {CCSN}  &  {\cite{N13papaer}}  \\ 
 {PISN Model } & {CCSN + DC + PISN} & { $40 ~ M_\odot <  M \leq 100~M_\odot$ } & {interpolation} & { } \\
 { } & {  } &{$100 ~ M_\odot <  M \leq 140~M_\odot$} & {DC}&  {--} \\ 
 {  } & { } &{$140 ~ M_\odot <  M \leq 300~M_\odot$}&  {PISN} & {\cite{2018ApJ...857..111T}} \\
  \hline
 { } & { } & { $9 ~ M_\odot \leq  M < 13~M_\odot$} & {extrapolation} & { } \\
 CCSN Model & CCSN &{$13 ~ M_\odot \leq  M \leq 40~M_\odot$} & {CCSN}  &  {\cite{N13papaer}}  \\ 
\hline
{ } & { } & { $9 ~ M_\odot \leq  M < 13~M_\odot$} & {extrapolation} & { } \\
 HN Model & {HN} & { $13 ~ M_\odot \leq  M \leq 40~M_\odot$}  & {HN} &  {\cite{2008ApJ...673.1014U}} \\ 
\enddata
\tablecomments{(1) Model name. (2) Type of Superenova. We assume that stars with masses between $100-140~N_\odot$ cause direct collapse (DC). (3) Mass range of supernova. (4) Yields  (5) Yields reference.}
\label{table_literaturemodel}
\end{deluxetable*}

In the PISN models, we adopt the mass range from 9 $M_\odot$ to 300 $M_\odot$. The lower limit of the mass range, $M_\odot$, approximately corresponds to the lower limit mass of core-collapse supernovae that is just beyond the mass of the electron-capture supernovae \citep{N13papaer}. The upper limit of the mass range, 300 $M_\odot$, is chosen, because the upper limit mass of PISNe is $\sim 300 M_\odot$.

We derive lifetimes of the stars as a function of masses from \cite{1998A&A...334..505P} and \cite{2018ApJ...857..111T}.
The ranges of time calculated by our models are from $10^{6.28}~\mathrm{yr}$, which is the lifetimes of 300 $M_\odot$ star, to $10^{7.52}~\mathrm{yr}$, which is the lifetimes of 9 $M_\odot$ star.
The stars cause supernova explosions after finishing their lifetimes.
Adding up the ejecta of supernovae on the basis of model yields, we calculate the abundance ratios of galaxies.
%
We assume that stars in the mass range of 9-100 $M_\odot$ cause CCSNe, while stars in the mass range of 140-300 $M_\odot$ undergo PISNe. We use the PISN \citep{2018ApJ...857..111T} and the CCSN yields \citep{N13papaer} that cover 140-300 $M_\odot$ and 13-40 $M_\odot$, respectively. We obtain the CCSN yields in 9-13 $M_\odot$ by the extrapolation of the CCSN yields (13-40 $M_\odot$). The CCSN yields in 40-100 $M_\odot$ are calculated by the interpolation of the CCSN and PISN yields. Stars in the mass range of 100-140 $M_\odot$ are assumed to collapse directly into black holes.

The CCSN model in the second line of Table \ref{table_literaturemodel} is the same as the PISN model, but for the mass range of 9-40 $M_\odot$ that is free from the PISN contributions. 
Similarly, the HN model in the third line of Table \ref{table_literaturemodel} is the same as the CCSN model, but for the HN yield of \cite{2002ApJ...578..855U} in 13-40 $M_\odot$.
We also develop another CCSN and HN models with the yields including the mixing and fallback parameters for comparison. We summarize the parameter sets for the yields in Table \ref{table_ourmodel}.
\cite{N13papaer} and \cite{2008ApJ...673.1014U} calculate the yields of the CCSNe and HNe using the same assumption as this paper, but with different mixing regions.

\begin{deluxetable*}{cccccc}[htp]
\tabletypesize{\scriptsize}
\tablewidth{0pt} 
\tablecaption{Models of our yields \label{tab:deluxesplit}}
\tablehead{\colhead{Model} & \colhead{Supernova}& \colhead{Progenitor Star} & \colhead{ Yields } & \colhead{Mixing region factor} & \colhead{Metallicity}
}
\colnumbers
 \startdata
 { } & { } & { $9 ~ M_\odot \leq  M < 13~M_\odot$} & {extrapolation} & {  } & {  } \\
 CCSN-MF Model& CCSN & $13 ~ M_\odot \leq  M \leq 40~M_\odot$   & { CCSN  } & $x = 0, 0.1, 0.2$ & 0, 0.004\\ 
\hline
 { } & { } & { $9 ~ M_\odot \leq  M < 13~M_\odot$} & {extrapolation} & {  } & {  } \\
 HN-MF Model& HN &$13 ~ M_\odot \leq  M \leq 40~M_\odot$ &  { HN } & $x = 0, 0.1, 0.2$ & 0, 0.004\\ 
 \enddata
\tablecomments{(1) Model name. (2) Type of Superenova. (3) Mass range of supernova. (4) Yields. (5) Mixing region factor value. (6) Metallicity.} 
\label{table_ourmodel}
\end{deluxetable*}
In addition to the models explained above, we add the Type Ia SN yields by the mass fraction of the ejecta because Type Ia SNe can eject rich iron. We take the yields of Type Ia SNe from \cite{1984ApJ...286..644N} and \cite{1999ApJS..125..439I}.
This corresponds to a scenario where progenitors of Type Ia SNe are produced during an earlier formation epoch. 
The effect of Type Ia SN enrichment is implemented by adding the Type Ia SN ejecta to the abundance ratios of each model's endpoint and increasing the proportion of Type Ia SNe to 10 \% in order to investigate the effect of Type Ia SN enrichment.

%
%
\section{Results and Discussion} \label{sec:results}
\subsection{Comparing the EMPGs with the Models}

\begin{figure*}[htp]
    \centering
    \includegraphics[width=15cm]{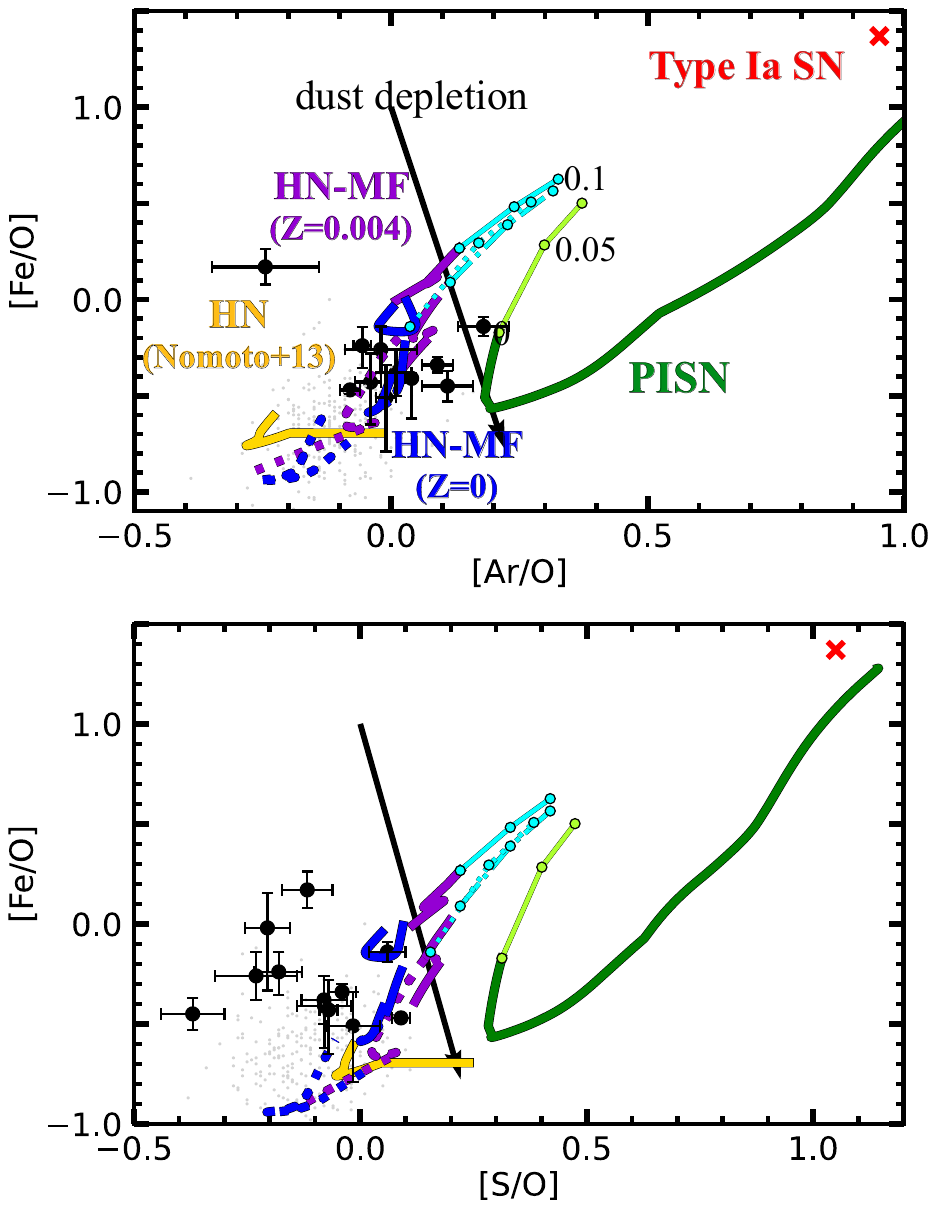}
    \label{fig:result_HN}
\end{figure*}
\begin{figure*}[htp]
    \centering
    \includegraphics[width=15cm]{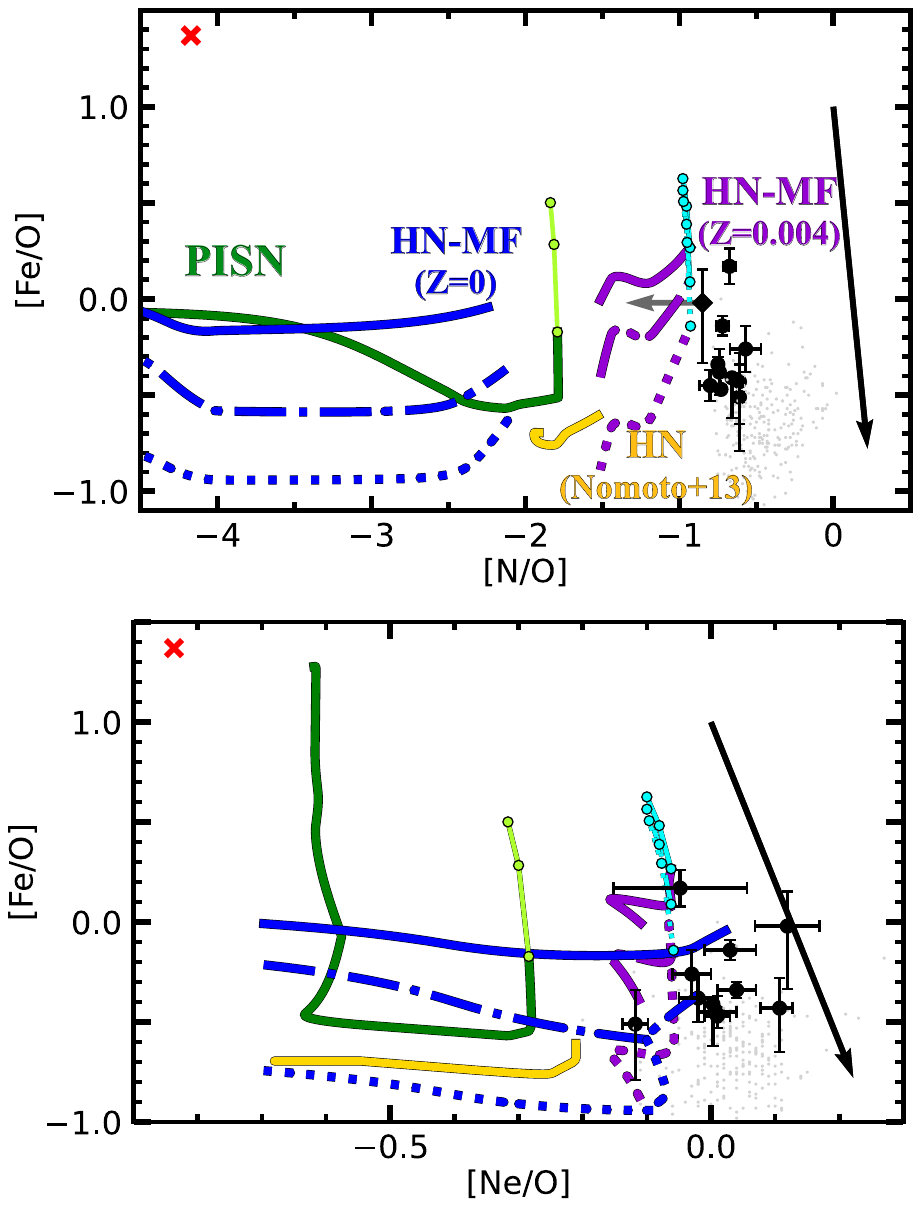}
    \caption{Comparisons of the EMPGs with the PISN and HN models in the abundance ratios. The top, second, third, and bottom panels present [Fe/O] as a function of [Ar/O], [S/O], [N/O], and [Ne/O], respectively.
    The black and gray circles show the EMPGs and the local dwarf galaxies \citep{2006A&A...448..955I}, respectively. 
    The green, purple, and blue lines present the time variation of PISN, the $Z=0.004$ HN-MF, and the $Z=0$ HN-MF models, respectively. 
    The yellow line shows the time variation of the HN model calculated with the yields of \cite{N13papaer}.
    The solid, dash-dotted, and dotted lines indicate $x=0$, $x=0.1$, and $x=0.2$ for the HN-MF models, respectively.  
    The red crosses are the abundance ratios of Type Ia SN ejecta \citep{1999ApJS..125..439I}.
    The light green and cyan curves represent 
    the PISN and $Z=0.004$ HN models with added the Type Ia SN ejecta.
    The numbers accompanied by the light green and light purple curves indicate the proportions of Type Ia SNe to HNe and PISNe, respectively.
    The effects of the dust depletion are denoted with the black arrows whose lengths present the change of the values from [0, 1.3] on the planes adapted from \cite{2013hbic.book.....F}.} 
    
    \label{fig:result_HN}
\end{figure*}
%
%
\begin{figure*}
    \centering
    \includegraphics[width=15cm]{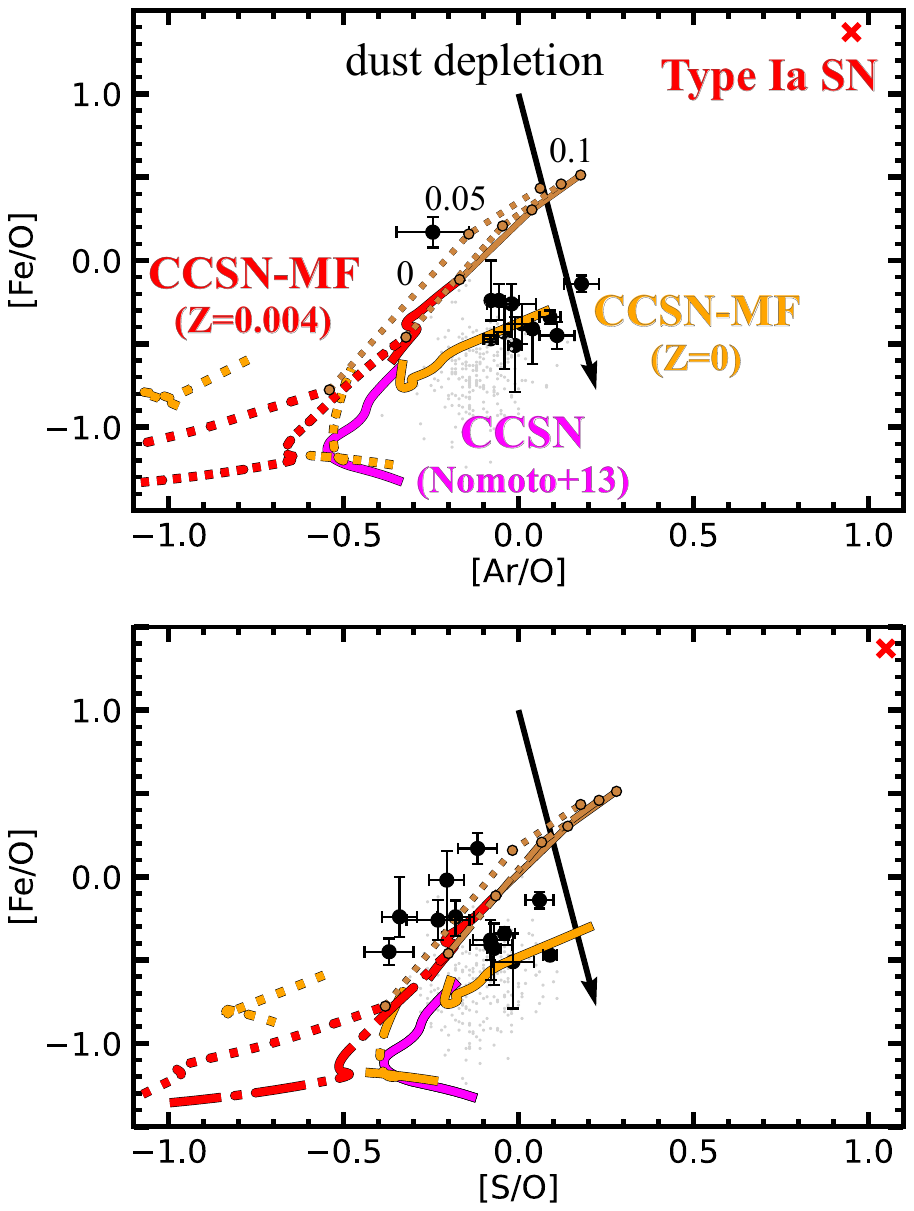}
    \label{fig:result_CCSN}
\end{figure*}

\begin{figure*}
    \centering
    \includegraphics[width=15cm]{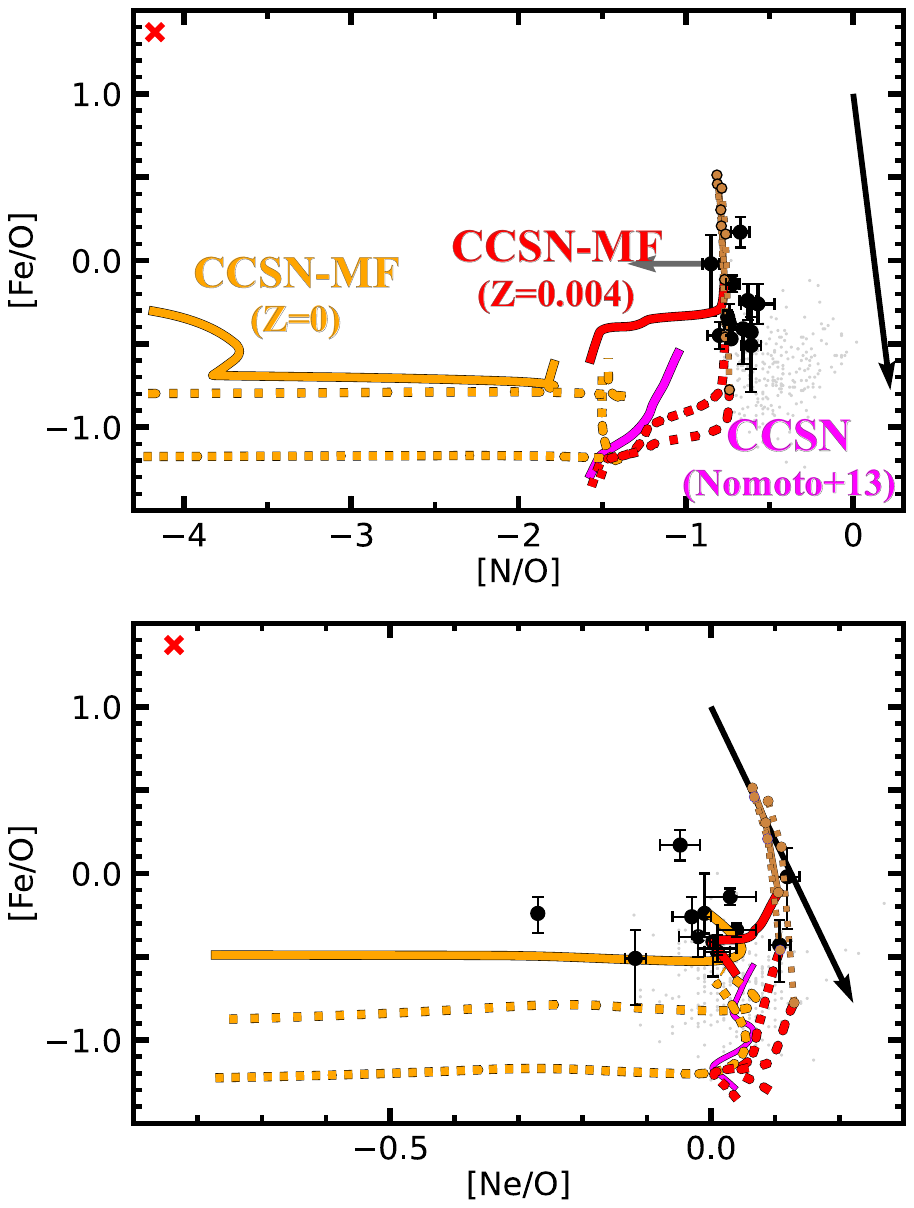}
    \caption{
    Same as Fig.\ref{fig:result_HN}, but for the CCSN models.
    The red and orange lines indicate the $Z=0.004$ CCSN-MF and $Z=0$ CCSN-MF models, respectively.
    The brown curves present $Z=0.004$ CCSN-MF models with added the Type Ia SN ejecta.
    The magenta curves show the CCSN model with the yields of \cite{N13papaer}.
    }
\end{figure*}

Figure \ref{fig:result_HN} compares the EMPGs (Section \ref{subsec:sample}) with the PISN (the green curves) and HN (the blue and purple curves) models (Section \ref{sec:model_develop})
on the [Fe/O] vs. [Ar/O], [S/O], [N/O], and [Ne/O] planes. 
Because of explosive O-burning, the [Ar/O] and [S/O] values are higher than CCSNe.
While the [Fe/O] values of the EMPGs are comparable with those of the PISN models, the [Ar/O] and [S/O] values of the EMPGs are significantly lower than those of the PISN models.
It is possible that the Ar and S abundances may be underestimated with the ICFs. However, the [S/O] and [Ar/O] abundance ratios in the EMPGs are lower than in the PISN models by $>0.5$ dex. Because the ICFs can increase the abundance ratios up to 0.5 dex even for EMPGs \citep{2006A&A...448..955I}, the low abundance ratios in the EMPGs are not explained by the ICF correction alone. 
For this reason, our conclusions do not change by the uncertainties of the ICF corrections.
The abundances of the EMPGs are generally higher than the local dwarf galaxies \citep{2006A&A...448..955I}.

The [N/O] and [Ne/O] values of the EMPGs are much higher than those of the PISN models.
The $x=0$ conditions of HNe can produce higher [Fe/O] values than the HN models of \cite{N13papaer} and lower [S/O] and [Ar/O] values than the PISN models.
The [Fe/O] values of the $x=0$ HN-MF models are comparable with the iron-rich([Fe/O] $>0$) EMPGs.
However, the [S/O] and [Ar/O] values of the $x=0$ HN-MF models are higher than the iron-rich EMPGs.
We conclude that the $x=0$ conditions of HN-MF models can not reproduce the iron-rich EMPGs.
In Figure \ref{fig:result_HN}, we calculate Spearman’s rank correlation coefficient to examine the potential correlation between [Fe/O] and [S/O], and find no statistically significant correlation between [Fe/O] and [S/O].
As mentioned in Section \ref{sec:abundance}, the abundance ratios for EMPGs are derived from gas-phase quantities, while our model predictions are based on total quantities, some of which can be depleted onto dust grains. 
To assess the potential impact of dust depletion on each abundance ratio diagram, we include arrows indicating the degree of depletion based on depletion factors defined by \cite{2013hbic.book.....F}, where depletion factors of 0.6 and $10^{-2}$ are assumed for O and Fe, respectively, and no depletion is considered for Ar, S, Ne, and N. 
For example, the value of 0.6 for O indicates that 60 \% of oxygen gas remains after dust depletion.

While we understand that the assumed values for dust depletions are still an open question, particularly in the low metallicity regime, we simply adopt these values in this paper to evaluate the possible effect of dust depletion and compare our models with the observations.
These differences between the EMPGs and PISN models in [Ar/O], [S/O], [N/O], and [Ne/O] cannot be explained by the effects of the dust depletion indicated with the arrows in Figure \ref{fig:result_HN}, suggesting that the chemical enrichment of the EMPGs is not dominated by PISNe. 
\cite{2021MNRAS.505.2361A} argue that the dust depletion of Ar is comparable to the one of oxygen, while S experiences almost no depletion. Using the dust depletions of \cite{2021MNRAS.505.2361A}, we confirm that the dust depletions little change the distribution of the EMPGs in Figures \ref{fig:result_HN} and \ref{fig:result_CCSN}. We think that the differences of dust depletion for S and Ar are not a major concern.
Similarly, Figure \ref{fig:result_HN} suggests none of the currently available HN models can fully explain all of the observed abundance ratios of EMPGs simultaneously.
While the [Fe/O] of the HN-MF models are as high as those of the EMPGs, the $Z=0$ models underpredict nitrogen and the $Z=0.004$ models overpredict sulfur.

Figure \ref{fig:result_CCSN} presents the CCSN models 
with two metallicities of $Z=0$ and $0.004$
for three mixing region factors of $x=0$, $0.1$, and $0.2$.
In Figure \ref{fig:result_CCSN}, the $Z=0.004$ models with $x=0.1$ and $0.2$ explain the iron-poor ([Fe/O] $\leq 0$) EMPGs. 
Similarly, these models agree with the abundance ratios of the local metal-poor galaxies (Section \ref{sec:MPS}). 
The iron-rich ([Fe/O] $>0$) EMPGs are not reproduced by these models, but the CCSN models with Type Ia SNe. 
The CCSN models with Type Ia show high [Fe/O] values without increasing the [Ar/O] and [S/O] values as much as the PISN models.
In Figure \ref{fig:result_CCSN}, a proportion of Type Ia SNe at $\sim 0.05$ can reproduce the abundance ratios of the iron-rich EMPGs. While the $x=0$ condition, which corresponds to the mixing and fallback mechanism being turned off, results in a [Fe/O] ratio higher than the $x=0.1$ and 0.2 conditions by $\sim 0.5$ dex, it is not necessarily required for the EMPGs, if an enrichment of Type Ia SNe is included.

\cite{2022ApJ...925..111I} conclude that the iron-rich EMPGs are not enriched by Type Ia SNe because the N/O values of EMPGs are lower than the chemical evolution models \citep{2016MNRAS.458.3466V,2018ApJ...852..101S}.
They assume the typical delay time of Type Ia SNe such as the MW.
These models show higher N/O and lower Fe/O values than the iron-rich EMPGs because the enrichment of Type Ia SNe and AGB stars is effective.
However, we introduce the Type Ia SNe enrichment without the time information by adding up to the CCSN models.
We can see the N/O and Fe/O values before the enrichment of AGB and Type Ia SNe become dominant.
We can find the possibility of a shorter Type Ia SNe delay time.

In summary, we conclude that the iron-rich EMPGs are not enriched by PISNe because the iron-rich EMPGs present lower [Ar/O] and [S/O] values than those of the PISN models.
The abundance ratios of the iron-poor EMPGs are explained by the $Z=0.004$ CCSN models with $x=0.1$ and $0.2$ in Figure \ref{fig:result_CCSN}.
We find no mixing \& fallback models of $x=0$ give higher [Fe/O] values than those of $x=0.1,0.2$ and the previous yields.
However, the HN or CCSN models alone can not reproduce the abundance ratios of the iron-rich EMPGs.
Although EMPGs are young galaxies, if Type Ia SNe occur after CCSNe, we can reproduce the abundance ratios of iron-rich EMPGs.

\subsection{
Comparing the EMPGs with Metal-Poor Stars
} \label{sec:MPS}

\begin{figure*}
    \centering
    \includegraphics[width=18cm]{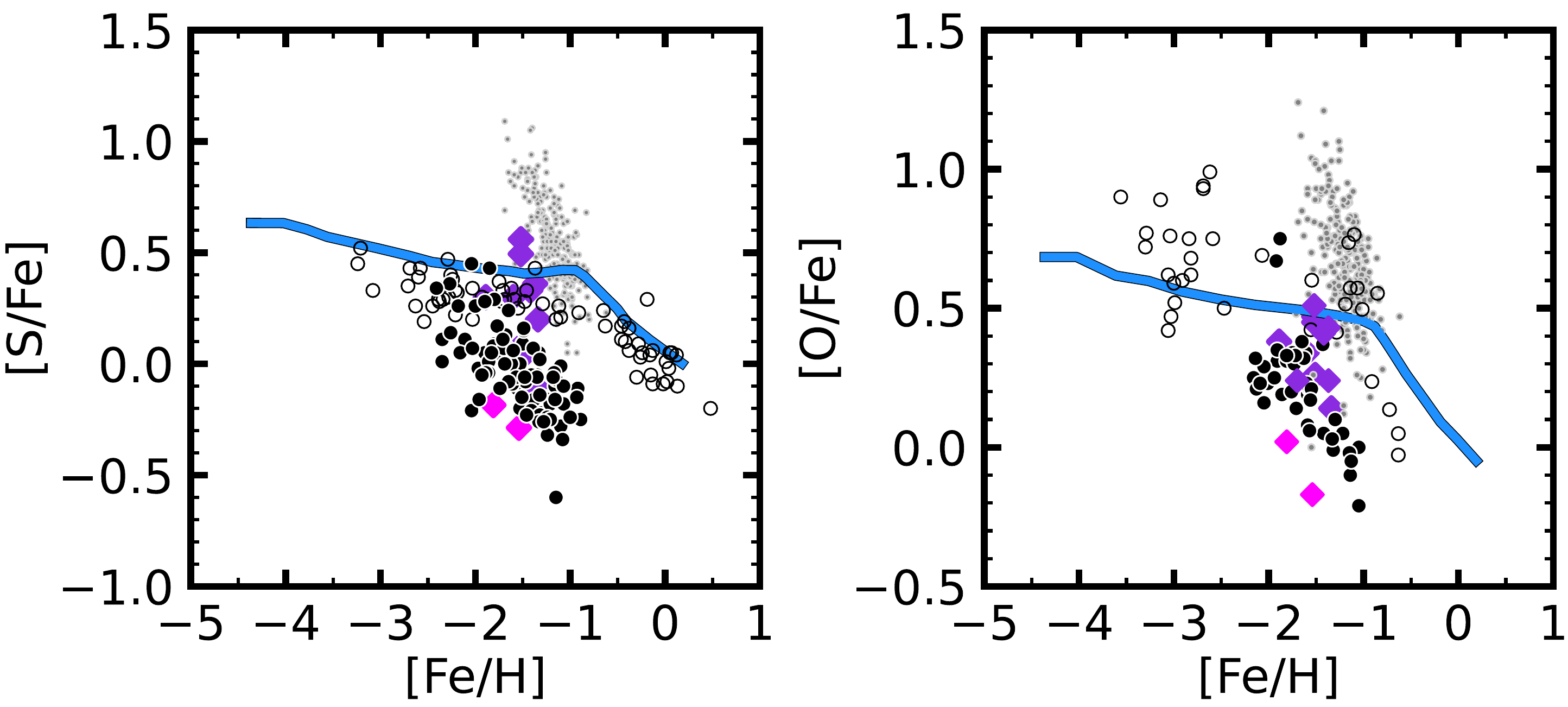}
    \caption{Comparisons of the EMPGs with metal-poor stars.
    The left and right panels present [S/Fe] and [O/Fe] as a function of [Fe/H], respectively.
    The magenta and purple diamonds show EMPGs with [Fe/O] $\geq0$ and [Fe/O] $<0$, respectively.
    The gray circles show the local dwarf galaxies \citep{2006A&A...448..955I}.
    The metal-poor stars in Milky Way are represented by open circles (\citealt{2002A&A...390..225C}, \citealt{2007A&A...469..319N}, \citealt{2004A&A...415..155B}, and \citealt{2004A&A...416.1117C}).
    The metal-poor stars in the Sculptor galaxy are shown by the black circles (\citealt{2015A&A...580A.129S} and \citealt{2023A&A...669A.125T}).
    The blue curves indicate the chemical evolution models \citep{N13papaer} including CCSNe, HNe, Type Ia SNe, AGB stars.}
    \label{fig:Fe_H}
\end{figure*}

 In Figure \ref{fig:Fe_H}, we compare the EMPGs with local dwarf galaxies and metal-poor stars in the MW and the Sculptor galaxy on the plots of [S/Fe] and [O/Fe] as a function of [Fe/H] that are useful to understand the chemical enrichment history. 
 The metal-poor stars in the Sculptor galaxy show low sulfur and high iron values, similar to the EMPGs. Here, we use the abundance ratio of [O/Fe] instead of [Fe/O], following the notation typically adopted in studies of metal-poor stars.
 We show the MW chemical evolution model of \citet[hereafter N13 model]{N13papaer} that includes CCSNe, HNe, Type Ia SNe, and AGB stars. Figure \ref{fig:Fe_H} indicates that the EMPGs with moderately small iron abundance ratios of [O/Fe]$ >0$ (purple diamonds) show abundance ratios similar to those of the metal-poor stars in the MW.
The rest of the EMPGs with [O/Fe]$ \le0$ (magenta diamonds) do not have the abundance ratios of the MW stars, but those of the Sculptor dwarf galaxy stars.
Oxygen is unstable due to the effects of explosions because oxygen is synthesized in a layer farther from Fe compared to Sulfur. The [O/Fe] values of metal-poor stars in Figure \ref{fig:Fe_H} exhibit greater scatter than the [S/Fe] values.
The [S/Fe] values do not scatter largely, because S and Fe are synthesized in the close layers. On the other hand, oxygen is synthesized in a layer far from Fe. These are reasons why the [O/Fe] values scatter largely than the [S/Fe] values.

In Figure \ref{fig:Fe_H}, we compare these observational data points with the chemical evolution tracks (blue curves) predicted by the N13 model, and confirm that the N13 model explains the abundance ratios of the MW stars.
In the N13 model, the iron abundance increases at around [Fe/H]$\gtrsim -1$ due to the chemical enrichment driven by Type-Ia SNe, which makes the knees at [Fe/H]$\sim -1$ in the chemical evolution tracks of [S/Fe] and [O/Fe] as a function of [Fe/H].

The observational data for [S/Fe] and [O/Fe] of the Sculptor galaxy stars (circles) show knees at [Fe/H]$\sim -2$ smaller than the one of the knees of the MW stars.
In the Sculptor galaxy, it is possible that the chemical enrichment of Type Ia SNe starts at [Fe/H]$\sim -2$ earlier than the MW galaxy of [Fe/H]$\sim -1$ (\citealt{2015A&A...580A.129S,2023A&A...669A.125T}).

\cite{2023A&A...669A.125T} claim that a minimum Type Ia SN delay time in the Sculptor galaxy is $10^8$ yr. Because the abundance ratios of the EMPGs with [O/Fe]$ \le0$ are similar to those of the Sculptor galaxy stars, the EMPGs may have a minimum Type Ia SN delay time as short as $10^8$ yr.
It is known that the delay time of a Type Ia SN in a galaxy including the MW is typically $\sim 10^{8.5}-10^9~$yr \citep{2021ApJ...922...15C}.
Type Ia SNe can theoretically occur in the delay time less than $10^8$ yr \citep{2009ApJ...699.2026R}.
\cite{2009ApJ...699.2026R} claim that the delay time depends on different evolutionary scenarios for Type Ia SNe.
\cite{2009ApJ...699.2026R} introduced three scenarios: the double degenerate scenario where two white dwarfs merge (\citealt{1984ApJS...54..335I,1984ApJ...277..355W}), the single degenerate scenario where a white dwarf accretes material from a hydrogen-rich companion star, and the AM Canum Venaticorum scenario where a white dwarf accretes material from a helium-rich companion star (\citealt{1982ApJ...253..798N, 2018SSRv..214...67N}).
Double degenerate and AM Canum Venaticirym scenarios present shorter delay time than H-rich single degenerate scenarios \citep{2009ApJ...699.2026R}.
The delay time of Type Ia SNe in the Sculptor galaxy and the EMPGs with [O/Fe]$ \le0$ may be shorter than the one of the MW.

In Figure \ref{fig:Fe_H}, we also compare the gas-phase abundance ratios of the local dwarf galaxies \citep{2006A&A...448..955I}. Interestingly, the local dwarf galaxies present the [S/Fe] and [O/Fe] values larger than those of the MW stars.
The abundance ratios of the local dwarf galaxies \citep{2006A&A...448..955I} and our EMPGs are derived using the same ICFs.
In the local dwarf galaxies, the [O/Fe] and [S/Fe] values are higher than EMPGs.
If the influence of ICFs is responsible for the elevated [O/Fe] and [S/Fe] values, the EMPGs should exhibit a similar trend. However, EMPGs do not exhibit a trend similar to that of the local dwarf galaxies.
We do not think that the abundance ratios of the local dwarf galaxies are not affected by the ICFs.
Instead, we think that this difference between the abundance ratios of the local dwarf galaxies and our EMPGs is due to the dust depletion. The local dwarf galaxies have higher metallicities compared to EMPGs, resulting in a higher dust content. The impact of dust depletion is greater in the local dwarf galaxies than in EMPGs.
Since our EMPGs samples have lower metallicities than the local dwarf galaxies \citep{2006A&A...448..955I}, the effects of dust depletion of our EMPGs are lower.

\subsection{Comparing JWST High-$z$ Galaxies with the EMPGs and the Models}

Recent observations with JWST have allowed us to directly measure abundance ratios including [S/O] and [Ar/O] for high-$z$ ($\gtrsim 4$) universe  (\citealt{2023arXiv230106696S, 2023arXiv230700710I}). 
It would be advantageous to conduct the above analysis using elemental abundances at higher redshifts, where previous star formation and subsequent Type Ia SN explosions are less complicating factors.

To investigate the elemental abundances in galaxies at $z\sim 4-10$ using JWST, we utilize 70 galaxies analyzed and presented in \cite{2023arXiv230112825N}. 
These spectra are taken from the three major public spectroscopic programs, including Early Release Observations \citep{2022ApJ...936L..14P}, GLASS \citep{2022ApJ...935..110T}, and CEERS \citep{2023ApJ...946L..13F}. 
\cite{2023arXiv230700710I} have calculated the abundance ratios of [S/O], [Ar/O], and [Ne/O] for these 70 galaxies of \citet{2023arXiv230112825N}
%
in the same manner as our methods.
The abundance ratios of 54 out of 70 galaxies can be determined or upper limits obtained.
For the other 16 galaxies, the upper limits of abundance ratios can not be measured.
We include the abundance ratios of 54 galaxies derived by \cite{2023arXiv230700710I} in our sample.

The stellar masses of these galaxies are $10^{7.5}-10^{9.5}~M_\odot$ \citep{2023arXiv230112825N}.
We can identify similarities and differences in terms of elemental abundances between high-$z$ galaxies and EMPGs. These comparisons, combined with our chemical evolution models, offer crucial insights into the formation mechanisms of EMPGs and young galaxies in the early universe.
Specifically, there is a question whether early young galaxies at high-$z$ are enriched by PISNe.
The PISN chemical enrichment is characterized by the rich iron 
via high [Fe/O] value, while the iron emission lines 
are too faint to be clearly detected in high-$z$ galaxies 
even by JWST deep spectroscopy with gravitational
lensing magnification \citep{2022A&A...666L...9C}.
It should be noted that our models suggest that, instead of [Fe/O], 
the [S/O] and [Ar/O] ratios of PISNe are significantly
higher than those of CCSNe+HNe (Figure \ref{fig:result_CCSN}).
Sulfur and Argon emission lines are moderately strong in
star-forming galaxies, and such lines are already detected
by JWST observations in various studies \citep{2023ApJ...944L..36P}.
Although [Fe/O] ratios cannot be investigated, 
we can test the possibility of
PISN chemical enrichment in the high redshift universe
with [S/O] and [Ar/O] ratios.
Strong Neon emission lines are also identified in these high-$z$ star-forming galaxies, and used for the abundance ratio 
studied \citep{2022ApJ...940L..23A}.

\begin{figure*}
    \centering
    \includegraphics[width=18cm]{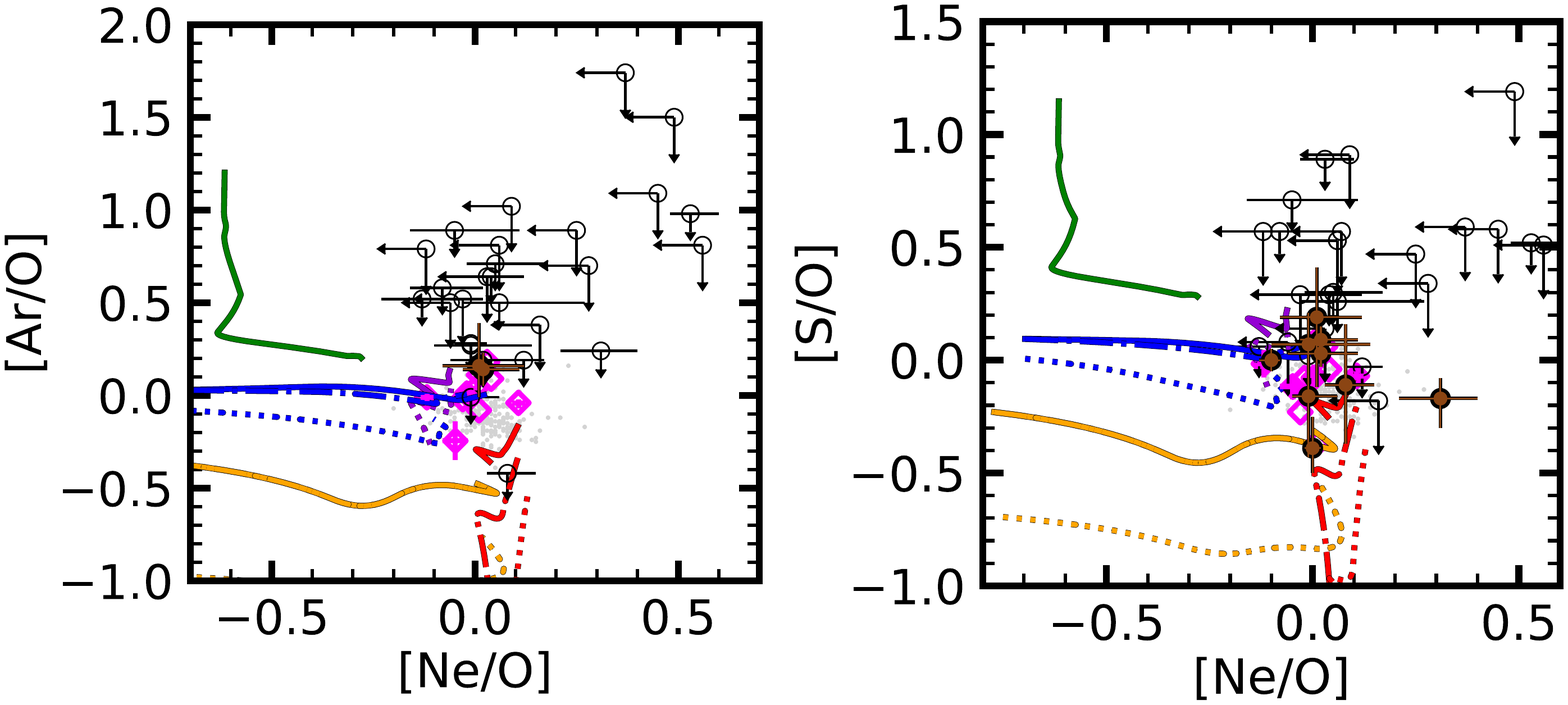}
    \caption{Comparison of JWST high-$z$ galaxies with EMPGs.
                The left and right panels present [Ar/O] and [S/O] as a function of [Ne/O], respectively. 
                The magenta diamonds show EMPGs.
                The filled brown and black open circles show JWST galaxies with measured abundance ratios and only upper limits, respectively.
                The color curves represent the chemical evolution models same as Fig \ref{fig:result_HN} and Fig \ref{fig:result_CCSN}.
                The gray circles show the local dwarf galaxies \citep{2006A&A...448..955I}.}
    \label{fig:JWST_result}
\end{figure*}

%
%
Figure \ref{fig:JWST_result} shows the abundance ratios of
[Ar/O] and [S/O] as a function of [Ne/O]
for the high-$z$ galaxies. Due to the limited signal-to-noise
ratios, there are only 2, and 10 out of 54
high-$z$ galaxies whose abundance ratios are determined
in the [Ar/O], and [S/O] vs. [Ne/O] planes, respectively.
The rest of the high-$z$ galaxies are shown with upper limits
in the abundance ratios.
In Figure \ref{fig:JWST_result},
we compare the high-$z$ galaxies with the local EMPGs and 
metal-poor galaxies investigated in Section \ref{sec:MPS}.
We find that the 2 and 10 high-$z$ galaxies with the abundance determinations have abundance ratios similar to those of 
the local EMPGs and metal-poor galaxies.
For the rest of the high-$z$ galaxies, the upper limits are
too weak to distinguish between the CCSNe, HNe, and PISNe models.
Although there are many high-$z$ star-forming galaxies
whose upper limits of the abundance ratios are consistent with the PISN models, no high-$z$ galaxies have abundance ratios clearly support the possibility of a PISN chemical enrichment.

\subsubsection{The Origin of Rich Nitrogen}
Recently, \citet{2023arXiv230207256B} report bright 
nitrogen emission N{\sc iv]} $\mathrm{\lambda}$1486 and N{\sc iii]} $\mathrm{\lambda}$1748 in a star-forming galaxy, dubbed GN-z11, at $z=10.6$ with a high quality JWST/NIRSpec spectrum.
\cite{2023arXiv230210142C} and \cite{2023arXiv230304179S} derive nitrogen-oxygen abundance ratio of [N/O] $\gtrsim 0.5$, greater than four times solar.
Generally, nitrogen in the ISM is enriched by the asymptotic giant branch (AGB) star or the rotation of a star.
This high [N/O] is difficult to be explained at $z=10.6$ because the timescale of nitrogen enrichment by AGB stars is longer than that of oxygen enrichment.
In contrast with AGB stars, intermediate-mass stars enrich nitrogen early in stellar evolution ($\sim$3~Myr) by rotation.
%
The rotation of stars induces the mixing of the outer hydrogen and helium layers, making the CNO cycle more active.
This process enhances nitrogen abundance on the stellar surface \citep{2018ApJS..237...13L}.

\cite{2023arXiv230210142C} argue that Wolf-Rayet (WR) stars would produce the rich nitrogen of GN-z11.
In WR stars, a large amount of nitrogen is ejected by the stellar wind.
On the other hand, \cite{2023arXiv230307955C} claim that the nucleosynthesis within super-massive stars (SMS) with $10^4~M_\odot$ can reproduce the abundance ratios of GN-z11.

\begin{figure*}
    \centering
    \includegraphics[width=15cm]{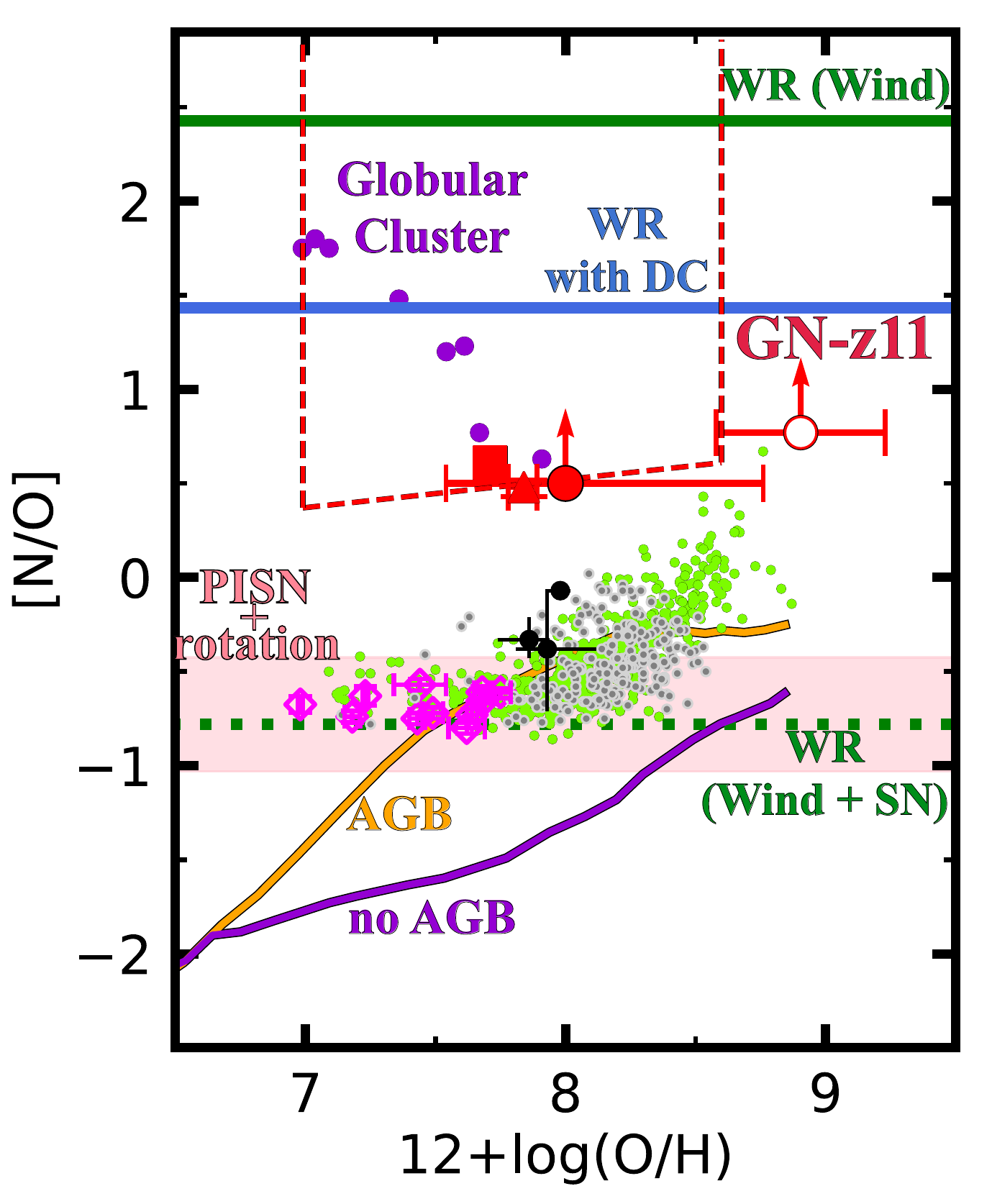}
    \caption{Comparison of GN-z11 with the models.
                The red dashed line defines the region indicating abundance ratios constraints of GN-z11 placed by \citet{2023arXiv230210142C}.
                The red square and triangle denote the abundance ratios of GN-z11 estimated by \cite{2023arXiv230304179S}.
                The red filled (open) circles show the abundance ratios of GN-z11 estimated by \cite{2023arXiv230700710I} in the case of stellar (AGN) photoionization.
                The purple circles represent dwarf turnoff stars in the globular cluster NGC 6752 \citep{2005A&A...433..597C}, while the green circles show $z\sim 0$ {\sc Hii} regions \citep{2012MNRAS.424.2316P,2020ApJ...893...96B}.
                The black circles present the JWST galaxies at $z=4-5$ \citep{2023arXiv230700710I}.
                The magenta diamonds and the gray circles denote the EMPGs and the local dwarf galaxies \citep{2006A&A...448..955I}, respectively.
                The orange (purple) curve indicates the chemical evolution model including CCSNe, HNe, and IaSNe with (no) AGB stars \citep{N13papaer}.
                The pink shade shows the range of abundance ratios for the rotating PISN model \citep{2018ApJ...857..111T}.
                The green solid (dash-dotted) line represents the abundance ratios of the ejecta from Wolf-Rayet stars via stellar winds (and CCSNe) \citep{2006A&A...447..623M}.
                The blue line indicates the abundance ratios for the direct collapse of massive WR stars \citep{2018ApJS..237...13L}.
                }
    \label{fig:N_O_result}
\end{figure*}

Figure \ref{fig:N_O_result} compares our galaxies and models with GN-z11 in $\log$(N/O) vs. $\log$(O/H), and explores the major physical origin of the rich nitrogen of GN-z11, following the previous studies \citep{2023arXiv230210142C,2023arXiv230304179S}.
\cite{2023arXiv230304179S} claim that the abundance ratio of GN-z11 is [N/O] $= 0.52^{+0.05}_{-0.04}$ to reproduce the observed spectral lines from the photoionization model, while
\cite{2023arXiv230700710I} obtain 
the [N/O]$>0.50$ and $12+\log$(O/H)$=8.00^{+0.76}_{-0.46}$ ([N/O] $>0.77$ and $12+\log$(O/H)$=8.58 - 9.23$) in the case of stellar (AGN) photoionization.
In Figure \ref{fig:N_O_result}, we show our EMPGs samples, JWST samples, the local dwarf galaxies \citep{2006A&A...448..955I}, the local {\sc Hii} regions \citep{2012MNRAS.424.2316P,2020ApJ...893...96B}, and the stars in the globular cluster 
NGC 6752 \citep{2005A&A...433..597C}
whose N/O values are as large as those of GN-z11 \citep{2023arXiv230304179S}.
%
We develop the rotating PISN models to consider the effect of the rotation.
The PISN models are made in the same manner as Sec \ref{sec:model_develop}, but with the rotating PISN yields obtained in \cite{2018ApJ...857..111T}.
We calculate the N/O values of the ejecta from the $60~M_\odot$ WR stars with stellar winds, referred to as "WR (Wind only)", and the stars with stellar winds and the subsequent CCSNe, referred to as "WR (Wind+SN)" \citep{2006A&A...447..623M}.
Because the rotating PISN models and the WR star models do not have values of 12+$\log$O/H, we show the region and lines of the [N/O] values for each model.
We use the N13 model that includes AGB stars.

The [N/O] values of GN-z11 and NGC 6752 are much higher than the other objects.
Our EMPGs and some of the local {\sc Hii} regions have lower oxygen abundance than GN-z11, while no other objects show [N/O] values as large as those of GN-z11 and NGC 6752.
Majority of local galaxies do not have N/O values as large as GN-z11 and NGC 6752, while N/O values of high-$z$ galaxies are poorly constrained except for GN-z11. There is a possibility that rich nitrogen galaxies may be more commonly found in the early universe.
The rotating PISN model and N13 model with AGB stars present richer [N/O] values than the N13 model with no AGB stars, while these [N/O] values are not as high as the one of GN-z11.

The WR (Wind only) model can produce high [N/O] values of [N/O]$\sim 2.43$.
As shown in WR (Wind only), once CCSNe take place, the [N/O] values decrease to $\sim -0.78$.
Because typical WR stars are massive stars and the stellar lifetime is short, it is difficult to keep the high N/O values for more than 7 Myr if they explode as CCSNe.
However, WR stars with a mass of $\gtrsim 25~M_\odot$ may directly collapse to black holes in low metallicity because they have massive Fe cores \citep{2020ApJ...888...91E}.

Here, in the same manner as Section \ref{sec:model_develop}, we develop a chemical evolution model with WR stars that directly collapse (hereafter WR-DC model).
We use the yields and lifetimes predicted by \citet{2018ApJS..237...13L} for rotating WR stars with direct collapse, such as formed in a low metallicity environment, in a mass range of $25-120~M_\odot$.
We adopt the yields of WR stars with metallicity [Fe/H] $= -3$ and rotation velocity $v = 300~\mathrm{km~s^{-1}}$.
We use the yields of rotating CCSNe with progenitor masses of $13-25~M_\odot$ derived in \cite{2018ApJS..237...13L}. 
In our WR-DC model, we assume that $25~M_\odot$ is a transition mass for direct collapse ($> 25~M_\odot$) and no-direct collapse ($\leq 25~M_\odot$) and that the stars with $25-120~M_\odot$ and $13-25~M_\odot$ end up as direct-collapse WR stars and CCSNe, respectively.
These are metal-poor WR stars whose cores are massive 
at the final stage of the stellar evolution \citep{2002RvMP...74.1015W,2020ApJ...888...91E}.
For this reason, \citet{2018ApJS..237...13L} assume that these WR stars directly collapse with no CCSN events, and that the WR stars only produce ejecta via stellar winds that are rich in nitrogen.

\begin{figure*}
    \centering
    \includegraphics[width=18.5cm]{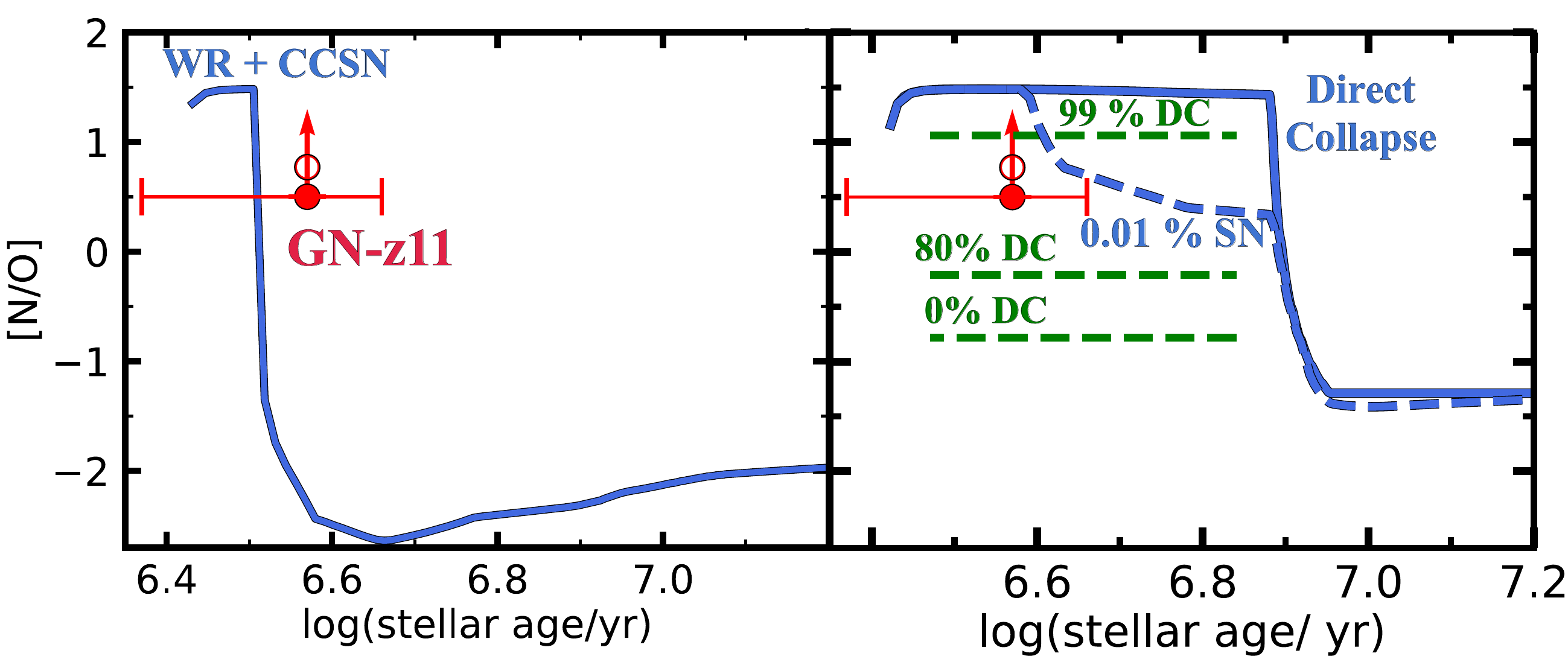}
    \caption{Comparison of GN-z11 with the WR-DC model. The red filled (open) circles show the abundance ratios of GN-z11 estimated by \cite{2023arXiv230700710I} in the case of stellar (AGN) photoionization.
    The left panel and right panels present the WR star models without direct collapse and with direct collapse, respectively.
    The blue curves show our WR-DC model developed with the direct-collapse WR stars \citep{2018ApJS..237...13L}.
    The green dashed lines represent the [N/O] values of varying the percentage of direct-collapse WR stars \citep{2006A&A...447..623M}, where the lengths of the green dashed lines indicate the lifetimes of stars in the mass range of $30-90~M_\odot$.
    }
    \label{fig:N_O_age}
\end{figure*}

Figure \ref{fig:N_O_age} presents GN-z11 and the WR-DC model in the plane of [N/O] vs. the stellar age.
In Figure \ref{fig:N_O_age}, the nitrogen abundance of the WR-DC model decreases rapidly at $\sim 10^{6.9}~\mathrm{yr}$ that corresponds to the lifetime of stars with the transition
mass of $25~M_\odot$.
Figure \ref{fig:N_O_age} indicates that the stellar age and nitrogen abundance of GN-z11 fall on the WR-DC model. The high nitrogen abundance of GN-z11 ([N/O]$\sim 0.5$) can be explained by the WR-DC model for the given young age ($3-5$ Myr).
Figure \ref{fig:N_O_age} also compares independent yield models of \cite{2006A&A...447..623M} that are presented in Figure \ref{fig:N_O_result}. Again, the yield of WR(Wind+SN) shows [N/O] as low as $-0.78$. Here, we assume that some fractions of WR stars directly collapse, and calculate [N/O] values for the fractions of direct collapse WR stars. With this yield model, We find that $\sim 97$\% of the WR stars need to directly collapse to produce [N/O] as high as the one of GN-z11 ([N/O]$\sim 0.52$). This high fraction of the direct-collapse WR stars is consistent with that of the WR-DC model whose WR stars with $>25 M_\odot$ fully directly collapse.

Note that the models of direct collapse WR stars assume no CCSN ejecta in the initial condition. In other words, these models predict [N/O] for an onset of initial star-formation from primordial gas. Although the models of direct collapse WR stars explain the high [N/O] value of GN-z11, it is unclear whether 
the moderately matured system of GN-z11 with stellar masses of $10^8-10^9M_\odot$ and [O/H]$\sim 0.1$ can be reached in the short time-scale at least within $\lesssim 10^{6.9}~\mathrm{yr}$ corresponding to the lifetime of $\gtrsim 25~M_\odot$ stars that could directly collapse. 
First for the stellar mass, GN-z11 has a high star-formation rate of $20 M_\odot$ yr$^{-1}$ \citep{2023arXiv230304179S}, it takes only $5-50$ Myr to produce $10^8-10^9M_\odot$ stars in the constant star-formation history. Because it is comparable with the short time-scale of $\lesssim 10^{6.9}~\mathrm{yr}$, the moderately high stellar masses of $10^8-10^9M_\odot$ do not contradict with the initial star-formation. Second for the oxygen abundance, the moderately high value of [O/H]$\sim 0.1$ cannot be accomplished with no CCSNe, if one assumes that the moderately high [O/H] is the average value for the large hydrogen-gas reservoir of the interstellar medium in the GN-z11 galaxy. However, if WR stars do not explode as CCSNe, metals are not well mixed in GN-z11. A small amount of metals produced by WR-star winds are confined in compact star-forming regions with a relatively small amount of hydrogen gas (the scenario similar to \citealt{2018ARA&A..56...83B}), and the moderately high value of [O/H]$\sim 0.1$ could be obtained by emission line diagnostics for ionized gas of the star-forming regions with the JWST data.

\section{Summary}\label{sec:summary}
We study the elemental abundance ratios of the local metal-poor galaxies, JWST high-$z$ galaxies, and Galactic/Sculptor stars taken from the literature with the chemical evolution models.
We conduct spectroscopic observations for the EMPGs, obtain the spectra for the three EMPGs, and measure line fluxes of hydrogen, oxygen, iron, argon, sulfur, neon, and nitrogen to estimate the abundance ratios.
n addition to these EMPGs, we use the abundance ratios of 11 galaxies whose Fe/O ratios are obtained in the literature.
We confirm that some of the EMPGs have excessive iron abundance ratios similar to or beyond the solar abundance ratio.
Developing the chemical evolution models 
for CCSNe, HNe, and PISNe to investigate the observed 
abundance ratios, we calculate the yields of CCSNe and HNe including the mixing \& fallback mechanism that enhances the iron abundance. 
The main results of this paper are summarized below.
\begin{itemize}
    \item Although the [Fe/O] values of the PISN models are as high as those of some EMPGs as claimed by \citet{2022ApJ...925..111I}, [Ar/O] and [S/O] values of the PISN models are significantly higher than those of the EMPGs. 
    Because PISNe have no compact remnants, fractions of S and Ar are higher than CCSNe.
    We thus conclude that the origin of the rich iron in our EMPGs sample is not a PISN.
    
    \item We compare the EMPGs with the CCSN models whose mixing factors are $x=0$, $0.1$, and $0.2$ to 
    increase the iron production. 
    Although the mixing \& fallback cases of 
    $x=0.1$ and $0.2$ enhance iron abundance by 
    the iron transfer from the core to the outer layers of a star,
    the abundance ratio of Fe/O does not increase, but rather decreases. This is because the amount of oxygen also increases in the ejecta by the mixing \& fallback process. In contrast with our expectations, the case of $x=0$ gives a [Fe/O] ratio higher than those of $x=0.1$ and $0.2$, due 
    to no iron fallback. 
    We thus find that the high [Fe/O] ratios cannot be explained by the difference of the mixing \& fallback mechanism.

    \item If an enrichment of Type Ia SNe is included, the CCSNe models can reproduce the high [Fe/O], low [Ar/O], and low [S/O] ratios of the EMPGs. 
    It is possible that the high [Fe/O] of the EMPGs are produced by the combination of CCSNe and Type Ia SNe that are produced by old low-mass stars produced in the early-generation star formation in the EMPGs.
    We compare the EMPGs with stars in the MW and Sculptor galaxies. We find that the gas-phase [Fe/O] and [Fe/H] values of the EMPGs do not agree with stellar values in the MW galaxy, but in the Sculptor galaxy whose
    Type Ia SNe are produced in the early star-formation phase
    explaining the [Fe/O] higher ([Fe/H] lower) than the MW galaxy.
    Since the high [Fe/O] and low [Fe/H] in the EMPGs are similar to Sculptor's stellar chemical abundance ratios, the EMPGs may produce Type Ia SNe in their early star-formation phases.

    \item We investigate the abundance ratios [Ar/O], [S/O], and [Ne/O], of high$-z$ ($z=4-9$) galaxies recently measured with JWST.
    Although the iron abundance ratio is not obtained for these high-$z$ galaxies due to the weak iron emission, 
    our models suggest that [Ar/O] and [S/O] values are useful to distinguish chemical enrichments of PISNe and the other types of SNe. The comparisons with [Ar/O] and [S/O] indicate that there are no high-$z$ galaxies, so far identified, showing [Ar/O] and [S/O] values as high as those of PISNe, while many of the high-$z$ galaxies have
    weak upper limits of [Ar/O] and [S/O] that leave possibilities of PISN-dominated 
    chemical enrichment. 

    \item  GN-z11 shows a high [N/O] ratio despite being at $z = 10.6$.
    The main candidates for nitrogen origin are WR stars and SMS stars \citep{2023arXiv230210142C,2023arXiv230307955C}.
    We compare the [N/O] values of GN-z11 with the ejecta from the rotating WR stars, the rotating PISNe, and the AGB stars. We find that the N/O values of the rotating PISN and AGB models do not reproduce the [N/O] value as high as the one of GN-z11.
    We calculate the [N/O] values of the gas ejected from the WR stars with the stellar winds and CCSNe \citep{2006A&A...447..623M}.
    The WR star (Wind only) shows [N/O] values much higher than GN-z11. However, once CCSNe take place, the [N/O] values fall below the one of GN-z11.
    We develop the chemical evolution model, dubbed WR-DC model, with WR stars using the yields and lifetimes of WR stars that directly collapse in the mass range of $25-120~M_\odot$ 
    \citep{2018ApJS..237...13L}.
    The high [N/O] value of GN-z11 is explained by the WR-DC model for the stellar age of GN-z11.
    With an independent model of a WR-star yield \citep{2006A&A...447..623M}, we also find that [N/O] of GN-z11 is reproduced, if $\sim 97$\% of the WR stars directly collapse.
    In this scenario with the direct-collapse WR stars, one needs to assume the initial star formation from primordial gas. 
    This assumption is consistent with the stellar mass of GN-z11 because the high star-formation rate of GN-z11 allows the production of the observed stellar masses within a comparable time scale for keeping a high [N/O] value in the WR-DC model.
    The high metallicity of GN-z11 ([O/H]$\sim 0.1$) is also explained if WR stars do not explode as CCSNe and the ejecta from the stellar winds are enclosed in compact star-forming regions with a small amount of hydrogen gas.

\end{itemize}

We thank Yusuke Shibuya, Hiroya Umeda, Akinori Matsumoto, Gen Chiaki, Hajime Fukushima, Hiroki Nagakura, Satoshi Kikuta, Shohei Aoyama, Hidenobu Yajima, and Yi Xu for giving us helpful comments.
This paper includes data gathered with 10 m Keck Telescope located at W. M. Keck Observatory, Hawaii which is operated as a scientific partnership among the California Institute of Technology, the University of California and the National Aeronautics and Space Administration. 
The Observatory was made possible by the generous financial support of the W. M. Keck Foundation.
The authors wish to recognize and acknowledge the very significant cultural role and reverence that the summit of Maunakea has always had within the indigenous Hawaiian community.  
We are most fortunate to have the opportunity to conduct observations from this mountain.
This work is based on observations made with the NASA/ESA/CSA James Webb Space Telescope. These observations are associated with programs 2736, 1324, and 1345. The authors acknowledge the ERO, GLASS, and CEERS teams led by Klaus M. Pontoppidan, Tommaso Treu, and Steven L. Finkelstein, respectively, for developing their observing programs with a zero-exclusive access period.

The Hyper Suprime-Cam (HSC) collaboration includes the astronomical communities of Japan and Taiwan, and Princeton University. The HSC instrumentation and software were developed by the National Astronomical Observatory of Japan (NAOJ), the Kavli Institute for the Physics and Mathematics of the Universe (Kavli IPMU), the University of Tokyo, the High Energy Accelerator Research Organization (KEK), the Academia Sinica Institute for Astronomy and Astrophysics in Taiwan (ASIAA), and Princeton University. Based on data collected at the Subaru Telescope and retrieved from the HSC data archive system, which is operated by the Subaru Telescope and Astronomy Data Center at NAOJ.

The Pan-STARRS1 Surveys (PS1) and the PS1 public science archive have been made possible through contributions by the Institute for Astronomy, the University of Hawaii, the Pan-STARRS Project Office, the Max-Planck Society and its participating institutes, the Max Planck Institute for Astronomy, Heidelberg and the Max Planck Institute for Extraterrestrial Physics, Garching, The Johns Hopkins University, Durham University, the University of Edinburgh, the Queen's University Belfast, the Harvard-Smithsonian Center for Astrophysics, the Las Cumbres Observatory Global Telescope Network Incorporated, the National Central University of Taiwan, the Space Telescope Science Institute, the National Aeronautics and Space Administration under Grant No. NNX08AR22G issued through the Planetary Science Division of the NASA Science Mission Directorate, the National Science Foundation Grant No. AST-1238877, the University of Maryland, Eotvos Lorand University (ELTE), the Los Alamos National Laboratory, and the Gordon and Betty Moore Foundation.

This paper is supported by World Premier International Research Center Initiative (WPI Initiative) and the joint research program of the Institute of Cosmic Ray Research (ICRR), the University of Tokyo.
This work is supported by KAKENHI(20H00180) Grant-in-Aid for Scientific Research (A) through the Japan Society for the Promotion of Science.
This research was supported by a grant from the Hayakawa Satio Fund awarded by the Astronomical Society of Japan.
This work has been supported in part by JSPS KAKENHI grant Nos. JP17K05382, JP20K04024, JP21H04499, JP23K03452 (K.Nomoto.).
Yuki Isobe, Masato Onodera, and Kimihiko Nakajima are supported by JSPS KAKENHI grant Nos. 21J20785, JP21K03622, and JP20K22373, respectively.
This work has been supported by JSPS Core-to-Core Program (grant number: JPJSCCA20210003).

This paper's English composition was partially enhanced through the use of ChatGPT (OpenAI 20209), although no sentences were created from scratch.

\bibliography{main}{}

\begin{thebibliography}{}
\expandafter\ifx\csname natexlab\endcsname\relax\def\natexlab#1{#1}\fi
\providecommand{\url}[1]{\href{#1}{#1}}
\providecommand{\dodoi}[1]{doi:~\href{http://doi.org/#1}{\nolinkurl{#1}}}
\providecommand{\doeprint}[1]{\href{http://ascl.net/#1}{\nolinkurl{http://ascl.net/#1}}}
\providecommand{\doarXiv}[1]{\href{https://arxiv.org/abs/#1}{\nolinkurl{https://arxiv.org/abs/#1}}}

\bibitem[{{Amayo} {et~al.}(2021){Amayo}, {Delgado-Inglada}, \& {Stasi{\'n}ska}}]{2021MNRAS.505.2361A}
{Amayo}, A., {Delgado-Inglada}, G., \& {Stasi{\'n}ska}, G. 2021, \mnras, 505, 2361, \dodoi{10.1093/mnras/stab1467}

\bibitem[{{Arellano-C{\'o}rdova} {et~al.}(2022){Arellano-C{\'o}rdova}, {Berg}, {Chisholm}, {Arrabal Haro}, {Dickinson}, {Finkelstein}, {Leclercq}, {Rogers}, {Simons}, {Skillman}, {Trump}, \& {Kartaltepe}}]{2022ApJ...940L..23A}
{Arellano-C{\'o}rdova}, K.~Z., {Berg}, D.~A., {Chisholm}, J., {et~al.} 2022, \apjl, 940, L23, \dodoi{10.3847/2041-8213/ac9ab2}

\bibitem[{{Asplund} {et~al.}(2021){Asplund}, {Amarsi}, \& {Grevesse}}]{Asplaud}
{Asplund}, M., {Amarsi}, A.~M., \& {Grevesse}, N. 2021, \aap, 653, A141, \dodoi{10.1051/0004-6361/202140445}

\bibitem[{{Bastian} \& {Lardo}(2018)}]{2018ARA&A..56...83B}
{Bastian}, N., \& {Lardo}, C. 2018, \araa, 56, 83, \dodoi{10.1146/annurev-astro-081817-051839}

\bibitem[{{Bensby} {et~al.}(2004){Bensby}, {Feltzing}, \& {Lundstr{\"o}m}}]{2004A&A...415..155B}
{Bensby}, T., {Feltzing}, S., \& {Lundstr{\"o}m}, I. 2004, \aap, 415, 155, \dodoi{10.1051/0004-6361:20031655}

\bibitem[{{Berg} {et~al.}(2021){Berg}, {Chisholm}, {Erb}, {Skillman}, {Pogge}, \& {Olivier}}]{2021ApJ...922..170B}
{Berg}, D.~A., {Chisholm}, J., {Erb}, D.~K., {et~al.} 2021, \apj, 922, 170, \dodoi{10.3847/1538-4357/ac141b}

\bibitem[{{Berg} {et~al.}(2019){Berg}, {Erb}, {Henry}, {Skillman}, \& {McQuinn}}]{2019ApJ...874...93B}
{Berg}, D.~A., {Erb}, D.~K., {Henry}, R. B.~C., {Skillman}, E.~D., \& {McQuinn}, K. B.~W. 2019, \apj, 874, 93, \dodoi{10.3847/1538-4357/ab020a}

\bibitem[{{Berg} {et~al.}(2020){Berg}, {Pogge}, {Skillman}, {Croxall}, {Moustakas}, {Rogers}, \& {Sun}}]{2020ApJ...893...96B}
{Berg}, D.~A., {Pogge}, R.~W., {Skillman}, E.~D., {et~al.} 2020, \apj, 893, 96, \dodoi{10.3847/1538-4357/ab7eab}

\bibitem[{{Brinchmann}(2023)}]{2023MNRAS.525.2087B}
{Brinchmann}, J. 2023, \mnras, 525, 2087, \dodoi{10.1093/mnras/stad1704}

\bibitem[{{Brocklehurst}(1971)}]{1971MNRAS.153..471B}
{Brocklehurst}, M. 1971, \mnras, 153, 471, \dodoi{10.1093/mnras/153.4.471}

\bibitem[{{Bunker} {et~al.}(2023){Bunker}, {Saxena}, {Cameron}, {Willott}, {Curtis-Lake}, {Jakobsen}, {Carniani}, {Smit}, {Maiolino}, {Witstok}, {Curti}, {D'Eugenio}, {Jones}, {Ferruit}, {Arribas}, {Charlot}, {Chevallard}, {Giardino}, {de Graaff}, {Looser}, {Luetzgendorf}, {Maseda}, {Rawle}, {Rix}, {Rodriguez Del Pino}, {Alberts}, {Egami}, {Eisenstein}, {Endsley}, {Hainline}, {Hausen}, {Johnson}, {Rieke}, {Rieke}, {Robertson}, {Shivaei}, {Stark}, {Sun}, {Tacchella}, {Tang}, {Williams}, {Willmer}, {Baker}, {Baum}, {Bhatawdekar}, {Bowler}, {Boyett}, {Chen}, {Circosta}, {Helton}, {Ji}, {Lyu}, {Nelson}, {Parlanti}, {Perna}, {Sandles}, {Scholtz}, {Suess}, {Topping}, {Uebler}, {Wallace}, \& {Whitler}}]{2023arXiv230207256B}
{Bunker}, A.~J., {Saxena}, A., {Cameron}, A.~J., {et~al.} 2023, arXiv e-prints, arXiv:2302.07256, \dodoi{10.48550/arXiv.2302.07256}

\bibitem[{{Calzetti} {et~al.}(2000){Calzetti}, {Armus}, {Bohlin}, {Kinney}, {Koornneef}, \& {Storchi-Bergmann}}]{2000ApJ...533..682C}
{Calzetti}, D., {Armus}, L., {Bohlin}, R.~C., {et~al.} 2000, \apj, 533, 682, \dodoi{10.1086/308692}

\bibitem[{{Cameron} {et~al.}(2023){Cameron}, {Katz}, {Rey}, \& {Saxena}}]{2023arXiv230210142C}
{Cameron}, A.~J., {Katz}, H., {Rey}, M.~P., \& {Saxena}, A. 2023, arXiv e-prints, arXiv:2302.10142, \dodoi{10.48550/arXiv.2302.10142}

\bibitem[{{Caminha} {et~al.}(2022){Caminha}, {Suyu}, {Mercurio}, {Brammer}, {Bergamini}, {Acebron}, \& {Vanzella}}]{2022A&A...666L...9C}
{Caminha}, G.~B., {Suyu}, S.~H., {Mercurio}, A., {et~al.} 2022, \aap, 666, L9, \dodoi{10.1051/0004-6361/202244517}

\bibitem[{{Cardelli} {et~al.}(1989){Cardelli}, {Clayton}, \& {Mathis}}]{1989ApJ...345..245C}
{Cardelli}, J.~A., {Clayton}, G.~C., \& {Mathis}, J.~S. 1989, \apj, 345, 245, \dodoi{10.1086/167900}

\bibitem[{{Carretta} {et~al.}(2005){Carretta}, {Gratton}, {Lucatello}, {Bragaglia}, \& {Bonifacio}}]{2005A&A...433..597C}
{Carretta}, E., {Gratton}, R.~G., {Lucatello}, S., {Bragaglia}, A., \& {Bonifacio}, P. 2005, \aap, 433, 597, \dodoi{10.1051/0004-6361:20041892}

\bibitem[{{Cayrel} {et~al.}(2004){Cayrel}, {Depagne}, {Spite}, {Hill}, {Spite}, {Fran{\c{c}}ois}, {Plez}, {Beers}, {Primas}, {Andersen}, {Barbuy}, {Bonifacio}, {Molaro}, \& {Nordstr{\"o}m}}]{2004A&A...416.1117C}
{Cayrel}, R., {Depagne}, E., {Spite}, M., {et~al.} 2004, \aap, 416, 1117, \dodoi{10.1051/0004-6361:20034074}

\bibitem[{{Charbonnel} {et~al.}(2023){Charbonnel}, {Schaerer}, {Prantzos}, {Ram{\'\i}rez-Galeano}, {Fragos}, {Kuruvandothi}, {Marques-Chaves}, \& {Gieles}}]{2023arXiv230307955C}
{Charbonnel}, C., {Schaerer}, D., {Prantzos}, N., {et~al.} 2023, arXiv e-prints, arXiv:2303.07955, \dodoi{10.48550/arXiv.2303.07955}

\bibitem[{{Chen} {et~al.}(2021){Chen}, {Hu}, \& {Wang}}]{2021ApJ...922...15C}
{Chen}, X., {Hu}, L., \& {Wang}, L. 2021, \apj, 922, 15, \dodoi{10.3847/1538-4357/ac178d}

\bibitem[{{Chen} {et~al.}(2002){Chen}, {Nissen}, {Zhao}, \& {Asplund}}]{2002A&A...390..225C}
{Chen}, Y.~Q., {Nissen}, P.~E., {Zhao}, G., \& {Asplund}, M. 2002, \aap, 390, 225, \dodoi{10.1051/0004-6361:20020735}

\bibitem[{{Curti} {et~al.}(2023){Curti}, {Maiolino}, {Carniani}, {D'Eugenio}, {Chevallard}, {Curtis-Lake}, {Looser}, {Scholtz}, {{\"U}bler}, {Witstok}, {Cameron}, {Charlot}, {Laseter}, {Sandles}, {Arribas}, {Bunker}, {Giardino}, {Maseda}, {Rawle}, {Rodr{\'\i}guez Del Pino}, {Smit}, {Willott}, {Eisenstein}, {Hausen}, {Johnson}, {Rieke}, {Robertson}, {Tacchella}, {Williams}, {Willmer}, {Baker}, {Bhatawdekar}, {Boyett}, {Egami}, {Helton}, {Ji}, {Kumari}, {Shivaei}, \& {Sun}}]{2023arXiv230408516C}
{Curti}, M., {Maiolino}, R., {Carniani}, S., {et~al.} 2023, arXiv e-prints, arXiv:2304.08516, \dodoi{10.48550/arXiv.2304.08516}

\bibitem[{{Ebinger} {et~al.}(2020){Ebinger}, {Curtis}, {Ghosh}, {Fr{\"o}hlich}, {Hempel}, {Perego}, {Liebend{\"o}rfer}, \& {Thielemann}}]{2020ApJ...888...91E}
{Ebinger}, K., {Curtis}, S., {Ghosh}, S., {et~al.} 2020, \apj, 888, 91, \dodoi{10.3847/1538-4357/ab5dcb}

\bibitem[{{Ferland}(2013)}]{2013hbic.book.....F}
{Ferland}, G.~J. 2013, {Hazy, A Brief Introduction to Cloudy C13.1}

\bibitem[{{Finkelstein} {et~al.}(2023){Finkelstein}, {Bagley}, {Ferguson}, {Wilkins}, {Kartaltepe}, {Papovich}, {Yung}, {Arrabal Haro}, {Behroozi}, {Dickinson}, {Kocevski}, {Koekemoer}, {Larson}, {Le Bail}, {Morales}, {P{\'e}rez-Gonz{\'a}lez}, {Burgarella}, {Dav{\'e}}, {Hirschmann}, {Somerville}, {Wuyts}, {Bromm}, {Casey}, {Fontana}, {Fujimoto}, {Gardner}, {Giavalisco}, {Grazian}, {Grogin}, {Hathi}, {Hutchison}, {Jha}, {Jogee}, {Kewley}, {Kirkpatrick}, {Long}, {Lotz}, {Pentericci}, {Pierel}, {Pirzkal}, {Ravindranath}, {Ryan}, {Trump}, {Yang}, {Bhatawdekar}, {Bisigello}, {Buat}, {Calabr{\`o}}, {Castellano}, {Cleri}, {Cooper}, {Croton}, {Daddi}, {Dekel}, {Elbaz}, {Franco}, {Gawiser}, {Holwerda}, {Huertas-Company}, {Jaskot}, {Leung}, {Lucas}, {Mobasher}, {Pandya}, {Tacchella}, {Weiner}, \& {Zavala}}]{2023ApJ...946L..13F}
{Finkelstein}, S.~L., {Bagley}, M.~B., {Ferguson}, H.~C., {et~al.} 2023, \apjl, 946, L13, \dodoi{10.3847/2041-8213/acade4}

\bibitem[{{Froese Fischer} \& {Tachiev}(2004)}]{FFT04}
{Froese Fischer}, C., \& {Tachiev}, G. 2004, Atomic Data and Nuclear Data Tables, 87, 1, \dodoi{10.1016/j.adt.2004.02.001}

\bibitem[{{Froese Fischer} {et~al.}(2006){Froese Fischer}, {Tachiev}, \& {Irimia}}]{atom_S3}
{Froese Fischer}, C., {Tachiev}, G., \& {Irimia}, A. 2006, Atomic Data and Nuclear Data Tables, 92, 607, \dodoi{10.1016/j.adt.2006.03.001}

\bibitem[{{Garnett}(1992)}]{1992AJ....103.1330G}
{Garnett}, D.~R. 1992, \aj, 103, 1330, \dodoi{10.1086/116146}

\bibitem[{{Hirschauer} {et~al.}(2016){Hirschauer}, {Salzer}, {Skillman}, {Berg}, {McQuinn}, {Cannon}, {Gordon}, {Haynes}, {Giovanelli}, {Adams}, {Janowiecki}, {Rhode}, {Pogge}, {Croxall}, \& {Aver}}]{2016ApJ...822..108H}
{Hirschauer}, A.~S., {Salzer}, J.~J., {Skillman}, E.~D., {et~al.} 2016, \apj, 822, 108, \dodoi{10.3847/0004-637X/822/2/108}

\bibitem[{{Iben} \& {Tutukov}(1984)}]{1984ApJS...54..335I}
{Iben}, I., J., \& {Tutukov}, A.~V. 1984, \apjs, 54, 335, \dodoi{10.1086/190932}

\bibitem[{{Ishigaki} {et~al.}(2018){Ishigaki}, {Tominaga}, {Kobayashi}, \& {Nomoto}}]{2018ApJ...857...46I}
{Ishigaki}, M.~N., {Tominaga}, N., {Kobayashi}, C., \& {Nomoto}, K. 2018, \apj, 857, 46, \dodoi{10.3847/1538-4357/aab3de}

\bibitem[{{Isobe} {et~al.}(2021){Isobe}, {Ouchi}, {Kojima}, {Shibuya}, {Hayashi}, {Rauch}, {Kikuchihara}, {Zhang}, {Ono}, {Fujimoto}, {Harikane}, {Kim}, {Komiyama}, {Kusakabe}, {Lee}, {Mawatari}, {Onodera}, {Sugahara}, \& {Yabe}}]{2021ApJ...918...54I}
{Isobe}, Y., {Ouchi}, M., {Kojima}, T., {et~al.} 2021, \apj, 918, 54, \dodoi{10.3847/1538-4357/ac05bf}

\bibitem[{{Isobe} {et~al.}(2022){Isobe}, {Ouchi}, {Suzuki}, {Moriya}, {Nakajima}, {Nomoto}, {Rauch}, {Harikane}, {Kojima}, {Ono}, {Fujimoto}, {Inoue}, {Kim}, {Komiyama}, {Kusakabe}, {Lee}, {Maseda}, {Matthee}, {Michel-Dansac}, {Nagao}, {Nanayakkara}, {Nishigaki}, {Onodera}, {Sugahara}, \& {Xu}}]{2022ApJ...925..111I}
{Isobe}, Y., {Ouchi}, M., {Suzuki}, A., {et~al.} 2022, \apj, 925, 111, \dodoi{10.3847/1538-4357/ac3509}

\bibitem[{{Isobe} {et~al.}(2023){Isobe}, {Ouchi}, {Tominaga}, {Watanabe}, {Nakajima}, {Umeda, H. R.}, {Yajima}, {Harikane}, {Fukushima}, {Xu}, {Ono}, \& {Zhang}}]{2023arXiv230700710I}
{Isobe}, Y., {Ouchi}, M., {Tominaga}, N., {et~al.} 2023, arXiv e-prints, arXiv:2307.00710, \dodoi{10.48550/arXiv.2307.00710}

\bibitem[{{Iwamoto} {et~al.}(1999){Iwamoto}, {Brachwitz}, {Nomoto}, {Kishimoto}, {Umeda, H. D.}, {Hix}, \& {Thielemann}}]{1999ApJS..125..439I}
{Iwamoto}, K., {Brachwitz}, F., {Nomoto}, K., {et~al.} 1999, \apjs, 125, 439, \dodoi{10.1086/313278}

\bibitem[{{Izotov} {et~al.}(2021){Izotov}, {Guseva}, {Fricke}, {Henkel}, {Schaerer}, \& {Thuan}}]{2021A&A...646A.138I}
{Izotov}, Y.~I., {Guseva}, N.~G., {Fricke}, K.~J., {et~al.} 2021, \aap, 646, A138, \dodoi{10.1051/0004-6361/202039772}

\bibitem[{{Izotov} {et~al.}(2009){Izotov}, {Guseva}, {Fricke}, \& {Papaderos}}]{2009A&A...503...61I}
{Izotov}, Y.~I., {Guseva}, N.~G., {Fricke}, K.~J., \& {Papaderos}, P. 2009, \aap, 503, 61, \dodoi{10.1051/0004-6361/200911965}

\bibitem[{{Izotov} {et~al.}(2006){Izotov}, {Stasi{\'n}ska}, {Meynet}, {Guseva}, \& {Thuan}}]{2006A&A...448..955I}
{Izotov}, Y.~I., {Stasi{\'n}ska}, G., {Meynet}, G., {Guseva}, N.~G., \& {Thuan}, T.~X. 2006, \aap, 448, 955, \dodoi{10.1051/0004-6361:20053763}

\bibitem[{{Izotov} \& {Thuan}(1998)}]{1998ApJ...497..227I}
{Izotov}, Y.~I., \& {Thuan}, T.~X. 1998, \apj, 497, 227, \dodoi{10.1086/305440}

\bibitem[{{Izotov} \& {Thuan}(1999)}]{1999ApJ...511..639I}
---. 1999, \apj, 511, 639, \dodoi{10.1086/306708}

\bibitem[{{Izotov} {et~al.}(2019){Izotov}, {Thuan}, \& {Guseva}}]{2019MNRAS.483.5491I}
{Izotov}, Y.~I., {Thuan}, T.~X., \& {Guseva}, N.~G. 2019, \mnras, 483, 5491, \dodoi{10.1093/mnras/sty3472}

\bibitem[{{Izotov} {et~al.}(2018){Izotov}, {Worseck}, {Schaerer}, {Guseva}, {Thuan}, {Fricke}, \& {Orlitov{\'a}}}]{2018MNRAS.478.4851I}
{Izotov}, Y.~I., {Worseck}, G., {Schaerer}, D., {et~al.} 2018, \mnras, 478, 4851, \dodoi{10.1093/mnras/sty1378}

\bibitem[{{Johansson} {et~al.}(2000){Johansson}, {Zethson}, {Hartman}, {Ekberg}, {Ishibashi}, {Davidson}, \& {Gull}}]{atom_fe_J}
{Johansson}, S., {Zethson}, T., {Hartman}, H., {et~al.} 2000, \aap, 361, 977

\bibitem[{{Kikuchihara} {et~al.}(2020){Kikuchihara}, {Ouchi}, {Ono}, {Mawatari}, {Chevallard}, {Harikane}, {Kojima}, {Oguri}, {Bruzual}, \& {Charlot}}]{2020ApJ...893...60K}
{Kikuchihara}, S., {Ouchi}, M., {Ono}, Y., {et~al.} 2020, \apj, 893, 60, \dodoi{10.3847/1538-4357/ab7dbe}

\bibitem[{{Kisielius} {et~al.}(2009){Kisielius}, {Storey}, {Ferland}, \& {Keenan}}]{atom_O2_coll}
{Kisielius}, R., {Storey}, P.~J., {Ferland}, G.~J., \& {Keenan}, F.~P. 2009, \mnras, 397, 903, \dodoi{10.1111/j.1365-2966.2009.14989.x}

\bibitem[{{Kojima} {et~al.}(2020){Kojima}, {Ouchi}, {Rauch}, {Ono}, {Nakajima}, {Isobe}, {Fujimoto}, {Harikane}, {Hashimoto}, {Hayashi}, {Komiyama}, {Kusakabe}, {Kim}, {Lee}, {Mukae}, {Nagao}, {Onodera}, {Shibuya}, {Sugahara}, {Umemura}, \& {Yabe}}]{2020ApJ...898..142K}
{Kojima}, T., {Ouchi}, M., {Rauch}, M., {et~al.} 2020, \apj, 898, 142, \dodoi{10.3847/1538-4357/aba047}

\bibitem[{{Kojima} {et~al.}(2021){Kojima}, {Ouchi}, {Rauch}, {Ono}, {Nakajima}, {Isobe}, {Fujimoto}, {Harikane}, {Hashimoto}, {Hayashi}, {Komiyama}, {Kusakabe}, {Kim}, {Lee}, {Mukae}, {Nagao}, {Onodera}, {Shibuya}, {Sugahara}, {Umemura}, \& {Yabe}}]{2021ApJ...913...22K}
---. 2021, \apj, 913, 22, \dodoi{10.3847/1538-4357/abec3d}

\bibitem[{{Kroupa}(2001)}]{2001MNRAS.322..231K}
{Kroupa}, P. 2001, \mnras, 322, 231, \dodoi{10.1046/j.1365-8711.2001.04022.x}

\bibitem[{{Limongi} \& {Chieffi}(2018)}]{2018ApJS..237...13L}
{Limongi}, M., \& {Chieffi}, A. 2018, \apjs, 237, 13, \dodoi{10.3847/1538-4365/aacb24}

\bibitem[{{Luridiana} {et~al.}(2015){Luridiana}, {Morisset}, \& {Shaw}}]{2015A&A...573A..42L}
{Luridiana}, V., {Morisset}, C., \& {Shaw}, R.~A. 2015, \aap, 573, A42, \dodoi{10.1051/0004-6361/201323152}

\bibitem[{{Mainali} {et~al.}(2018){Mainali}, {Zitrin}, {Stark}, {Ellis}, {Richard}, {Tang}, {Laporte}, {Oesch}, \& {McGreer}}]{2018MNRAS.479.1180M}
{Mainali}, R., {Zitrin}, A., {Stark}, D.~P., {et~al.} 2018, \mnras, 479, 1180, \dodoi{10.1093/mnras/sty1640}

\bibitem[{{Matsumoto} {et~al.}(2022){Matsumoto}, {Ouchi}, {Nakajima}, {Kawasaki}, {Murai}, {Motohara}, {Harikane}, {Ono}, {Kushibiki}, {Koyama}, {Aoyama}, {Konishi}, {Takahashi}, {Isobe}, {Umeda, H. R.}, {Sugahara}, {Onodera}, {Nagamine}, {Kusakabe}, {Hirai}, {Moriya}, {Shibuya}, {Komiyama}, {Fukushima}, {Fujimoto}, {Hattori}, {Hayashi}, {Inoue}, {Kikuchihara}, {Kojima}, {Koyama}, {Lee}, {Mawatari}, {Miyata}, {Nagao}, {Ozaki}, {Rauch}, {Saito}, {Suzuki}, {Takeuchi}, {Umemura}, {Xu}, {Yabe}, {Zhang}, \& {Yoshii}}]{2022ApJ...941..167M}
{Matsumoto}, A., {Ouchi}, M., {Nakajima}, K., {et~al.} 2022, \apj, 941, 167, \dodoi{10.3847/1538-4357/ac9ea1}

\bibitem[{{McLaughlin} \& {Bell}(2000)}]{atom_Ne_coll}
{McLaughlin}, B.~M., \& {Bell}, K.~L. 2000, Journal of Physics B Atomic Molecular Physics, 33, 597, \dodoi{10.1088/0953-4075/33/4/301}

\bibitem[{{Meynet} {et~al.}(2006){Meynet}, {Ekstr{\"o}m}, \& {Maeder}}]{2006A&A...447..623M}
{Meynet}, G., {Ekstr{\"o}m}, S., \& {Maeder}, A. 2006, \aap, 447, 623, \dodoi{10.1051/0004-6361:20053070}

\bibitem[{{Munoz Burgos} {et~al.}(2009){Munoz Burgos}, {Loch}, {Ballance}, \& {Boivin}}]{atom_ar}
{Munoz Burgos}, J.~M., {Loch}, S.~D., {Ballance}, C.~P., \& {Boivin}, R.~F. 2009, \aap, 500, 1253, \dodoi{10.1051/0004-6361/200911743}

\bibitem[{{Nakajima} {et~al.}(2023){Nakajima}, {Ouchi}, {Isobe}, {Harikane}, {Zhang}, {Ono}, {Umeda, H. R.}, \& {Oguri}}]{2023arXiv230112825N}
{Nakajima}, K., {Ouchi}, M., {Isobe}, Y., {et~al.} 2023, arXiv e-prints, arXiv:2301.12825, \dodoi{10.48550/arXiv.2301.12825}

\bibitem[{{Nishigaki} {et~al.}(2023){Nishigaki}, {Ouchi}, {Nakajima}, {Ono}, {Rauch}, {Isobe}, {Harikane}, {Narita}, {Zahedy}, {Xu}, {Yajima}, {Fukushima}, {Hirai}, {Kim}, {Inoue}, {Kusakabe}, {Lee}, {Nagao}, \& {Onodera}}]{2023arXiv230203158N}
{Nishigaki}, M., {Ouchi}, M., {Nakajima}, K., {et~al.} 2023, arXiv e-prints, arXiv:2302.03158, \dodoi{10.48550/arXiv.2302.03158}

\bibitem[{{Nissen} {et~al.}(2007){Nissen}, {Akerman}, {Asplund}, {Fabbian}, {Kerber}, {Kaufl}, \& {Pettini}}]{2007A&A...469..319N}
{Nissen}, P.~E., {Akerman}, C., {Asplund}, M., {et~al.} 2007, \aap, 469, 319, \dodoi{10.1051/0004-6361:20077344}

\bibitem[{{Nomoto}(1982)}]{1982ApJ...253..798N}
{Nomoto}, K. 1982, \apj, 253, 798, \dodoi{10.1086/159682}

\bibitem[{{Nomoto} {et~al.}(2013){Nomoto}, {Kobayashi}, \& {Tominaga}}]{N13papaer}
{Nomoto}, K., {Kobayashi}, C., \& {Tominaga}, N. 2013, \araa, 51, 457, \dodoi{10.1146/annurev-astro-082812-140956}

\bibitem[{{Nomoto} \& {Leung}(2018)}]{2018SSRv..214...67N}
{Nomoto}, K., \& {Leung}, S.-C. 2018, \ssr, 214, 67, \dodoi{10.1007/s11214-018-0499-0}

\bibitem[{{Nomoto} {et~al.}(1984){Nomoto}, {Thielemann}, \& {Yokoi}}]{1984ApJ...286..644N}
{Nomoto}, K., {Thielemann}, F.~K., \& {Yokoi}, K. 1984, \apj, 286, 644, \dodoi{10.1086/162639}

\bibitem[{{Peng} {et~al.}(2023){Peng}, {Vishwas}, {Stacey}, {Nikola}, {Lamarche}, {Rooney}, {Ball}, {Ferkinhoff}, \& {Spoon}}]{2023ApJ...944L..36P}
{Peng}, B., {Vishwas}, A., {Stacey}, G., {et~al.} 2023, \apjl, 944, L36, \dodoi{10.3847/2041-8213/acb59c}

\bibitem[{{Pilyugin} {et~al.}(2012){Pilyugin}, {Grebel}, \& {Mattsson}}]{2012MNRAS.424.2316P}
{Pilyugin}, L.~S., {Grebel}, E.~K., \& {Mattsson}, L. 2012, \mnras, 424, 2316, \dodoi{10.1111/j.1365-2966.2012.21398.x}

\bibitem[{{Pontoppidan} {et~al.}(2022){Pontoppidan}, {Barrientes}, {Blome}, {Braun}, {Brown}, {Carruthers}, {Coe}, {DePasquale}, {Espinoza}, {Marin}, {Gordon}, {Henry}, {Hustak}, {James}, {Jenkins}, {Koekemoer}, {LaMassa}, {Law}, {Lockwood}, {Moro-Martin}, {Mullally}, {Pagan}, {Player}, {Proffitt}, {Pulliam}, {Ramsay}, {Ravindranath}, {Reid}, {Robberto}, {Sabbi}, {Ubeda}, {Balogh}, {Flanagan}, {Gardner}, {Hasan}, {Meinke}, \& {Nota}}]{2022ApJ...936L..14P}
{Pontoppidan}, K.~M., {Barrientes}, J., {Blome}, C., {et~al.} 2022, \apjl, 936, L14, \dodoi{10.3847/2041-8213/ac8a4e}

\bibitem[{{Portinari} {et~al.}(1998){Portinari}, {Chiosi}, \& {Bressan}}]{1998A&A...334..505P}
{Portinari}, L., {Chiosi}, C., \& {Bressan}, A. 1998, \aap, 334, 505.
\newblock \doarXiv{astro-ph/9711337}

\bibitem[{{Quinet}(1996)}]{atom_fe_Q}
{Quinet}, P. 1996, \aaps, 116, 573

\bibitem[{{Ramsbottom} \& {Bell}(1997)}]{atom_Ar4_coll}
{Ramsbottom}, C.~A., \& {Bell}, K.~L. 1997, Atomic Data and Nuclear Data Tables, 66, 65, \dodoi{10.1006/adnd.1997.0741}

\bibitem[{{Rhoads} {et~al.}(2023){Rhoads}, {Wold}, {Harish}, {Kim}, {Pharo}, {Malhotra}, {Gabrielpillai}, {Jiang}, \& {Yang}}]{2023ApJ...942L..14R}
{Rhoads}, J.~E., {Wold}, I. G.~B., {Harish}, S., {et~al.} 2023, \apjl, 942, L14, \dodoi{10.3847/2041-8213/acaaaf}

\bibitem[{{Ruiter} {et~al.}(2009){Ruiter}, {Belczynski}, \& {Fryer}}]{2009ApJ...699.2026R}
{Ruiter}, A.~J., {Belczynski}, K., \& {Fryer}, C. 2009, \apj, 699, 2026, \dodoi{10.1088/0004-637X/699/2/2026}

\bibitem[{{Rynkun} {et~al.}(2019){Rynkun}, {Gaigalas}, \& {J{\"o}nsson}}]{atom_RGJ19}
{Rynkun}, P., {Gaigalas}, G., \& {J{\"o}nsson}, P. 2019, \aap, 623, A155, \dodoi{10.1051/0004-6361/201834931}

\bibitem[{{Sanders} {et~al.}(2023){Sanders}, {Shapley}, {Topping}, {Reddy}, \& {Brammer}}]{2023arXiv230106696S}
{Sanders}, R.~L., {Shapley}, A.~E., {Topping}, M.~W., {Reddy}, N.~A., \& {Brammer}, G.~B. 2023, arXiv e-prints, arXiv:2301.06696, \dodoi{10.48550/arXiv.2301.06696}

\bibitem[{{Schaerer} {et~al.}(2022){Schaerer}, {Marques-Chaves}, {Barrufet}, {Oesch}, {Izotov}, {Naidu}, {Guseva}, \& {Brammer}}]{2022A&A...665L...4S}
{Schaerer}, D., {Marques-Chaves}, R., {Barrufet}, L., {et~al.} 2022, \aap, 665, L4, \dodoi{10.1051/0004-6361/202244556}

\bibitem[{{Schlafly} \& {Finkbeiner}(2011)}]{2011ApJ...737..103S}
{Schlafly}, E.~F., \& {Finkbeiner}, D.~P. 2011, \apj, 737, 103, \dodoi{10.1088/0004-637X/737/2/103}

\bibitem[{{Senchyna} {et~al.}(2023){Senchyna}, {Plat}, {Stark}, \& {Rudie}}]{2023arXiv230304179S}
{Senchyna}, P., {Plat}, A., {Stark}, D.~P., \& {Rudie}, G.~C. 2023, arXiv e-prints, arXiv:2303.04179, \dodoi{10.48550/arXiv.2303.04179}

\bibitem[{{Sk{\'u}lad{\'o}ttir} {et~al.}(2015){Sk{\'u}lad{\'o}ttir}, {Andrievsky}, {Tolstoy}, {Hill}, {Salvadori}, {Korotin}, \& {Pettini}}]{2015A&A...580A.129S}
{Sk{\'u}lad{\'o}ttir}, {\'A}., {Andrievsky}, S.~M., {Tolstoy}, E., {et~al.} 2015, \aap, 580, A129, \dodoi{10.1051/0004-6361/201525956}

\bibitem[{{Stark}(2016)}]{2016ARA&A..54..761S}
{Stark}, D.~P. 2016, \araa, 54, 761, \dodoi{10.1146/annurev-astro-081915-023417}

\bibitem[{{Stark} {et~al.}(2014){Stark}, {Richard}, {Siana}, {Charlot}, {Freeman}, {Gutkin}, {Wofford}, {Robertson}, {Amanullah}, {Watson}, \& {Milvang-Jensen}}]{2014MNRAS.445.3200S}
{Stark}, D.~P., {Richard}, J., {Siana}, B., {et~al.} 2014, \mnras, 445, 3200, \dodoi{10.1093/mnras/stu1618}

\bibitem[{{Storey} \& {Hummer}(1995)}]{atom_H0}
{Storey}, P.~J., \& {Hummer}, D.~G. 1995, \mnras, 272, 41, \dodoi{10.1093/mnras/272.1.41}

\bibitem[{{Storey} {et~al.}(2014){Storey}, {Sochi}, \& {Badnell}}]{atom_O3_coll}
{Storey}, P.~J., {Sochi}, T., \& {Badnell}, N.~R. 2014, \mnras, 441, 3028, \dodoi{10.1093/mnras/stu777}

\bibitem[{{Suzuki} \& {Maeda}(2018)}]{2018ApJ...852..101S}
{Suzuki}, A., \& {Maeda}, K. 2018, \apj, 852, 101, \dodoi{10.3847/1538-4357/aaa024}

\bibitem[{{Takahashi} {et~al.}(2018){Takahashi}, {Yoshida}, \& {Umeda}}]{2018ApJ...857..111T}
{Takahashi}, K., {Yoshida}, T., \& {Umeda}, H. 2018, \apj, 857, 111, \dodoi{10.3847/1538-4357/aab95f}

\bibitem[{{Tang} {et~al.}(2023){Tang}, {Zhang}, {Yan}, {Zhang}, {Carigi}, \& {Fern{\'a}ndez-Trincado}}]{2023A&A...669A.125T}
{Tang}, B., {Zhang}, J., {Yan}, Z., {et~al.} 2023, \aap, 669, A125, \dodoi{10.1051/0004-6361/202244052}

\bibitem[{{Tayal}(2011)}]{atom_N_coll}
{Tayal}, S.~S. 2011, \apjs, 195, 12, \dodoi{10.1088/0067-0049/195/2/12}

\bibitem[{{Tayal} \& {Gupta}(1999)}]{atom_S3_coll}
{Tayal}, S.~S., \& {Gupta}, G.~P. 1999, \apj, 526, 544, \dodoi{10.1086/307971}

\bibitem[{{Tayal} \& {Zatsarinny}(2010)}]{atom_S2_coll}
{Tayal}, S.~S., \& {Zatsarinny}, O. 2010, \apjs, 188, 32, \dodoi{10.1088/0067-0049/188/1/32}

\bibitem[{{Tominaga} {et~al.}(2007){Tominaga}, {Umeda, H. D.}, \& {Nomoto}}]{2007ApJ...660..516T}
{Tominaga}, N., {Umeda, H. D.}, \& {Nomoto}, K. 2007, \apj, 660, 516, \dodoi{10.1086/513063}

\bibitem[{{Treu} {et~al.}(2022){Treu}, {Roberts-Borsani}, {Bradac}, {Brammer}, {Fontana}, {Henry}, {Mason}, {Morishita}, {Pentericci}, {Wang}, {Acebron}, {Bagley}, {Bergamini}, {Belfiori}, {Bonchi}, {Boyett}, {Boutsia}, {Calabr{\'o}}, {Caminha}, {Castellano}, {Dressler}, {Glazebrook}, {Grillo}, {Jacobs}, {Jones}, {Kelly}, {Leethochawalit}, {Malkan}, {Marchesini}, {Mascia}, {Mercurio}, {Merlin}, {Nanayakkara}, {Nonino}, {Paris}, {Poggianti}, {Rosati}, {Santini}, {Scarlata}, {Shipley}, {Strait}, {Trenti}, {Tubthong}, {Vanzella}, {Vulcani}, \& {Yang}}]{2022ApJ...935..110T}
{Treu}, T., {Roberts-Borsani}, G., {Bradac}, M., {et~al.} 2022, \apj, 935, 110, \dodoi{10.3847/1538-4357/ac8158}

\bibitem[{{Trump} {et~al.}(2023){Trump}, {Arrabal Haro}, {Simons}, {Backhaus}, {Amor{\'\i}n}, {Dickinson}, {Fern{\'a}ndez}, {Papovich}, {Nicholls}, {Kewley}, {Brunker}, {Salzer}, {Wilkins}, {Almaini}, {Bagley}, {Berg}, {Bhatawdekar}, {Bisigello}, {Buat}, {Burgarella}, {Calabr{\`o}}, {Casey}, {Ciesla}, {Cleri}, {Cole}, {Cooper}, {Cooray}, {Costantin}, {Croton}, {Ferguson}, {Finkelstein}, {Fujimoto}, {Gardner}, {Gawiser}, {Giavalisco}, {Grazian}, {Grogin}, {Hathi}, {Hirschmann}, {Holwerda}, {Huertas-Company}, {Hutchison}, {Jogee}, {Juneau}, {Jung}, {Kartaltepe}, {Kirkpatrick}, {Kocevski}, {Koekemoer}, {Lotz}, {Lucas}, {Magnelli}, {Matharu}, {P{\'e}rez-Gonz{\'a}lez}, {Pirzkal}, {Rafelski}, {Rose}, {Seill{\'e}}, {Somerville}, {Straughn}, {Tacchella}, {Vanderhoof}, {Weiner}, {Wuyts}, {Yung}, \& {Zavala}}]{2023ApJ...945...35T}
{Trump}, J.~R., {Arrabal Haro}, P., {Simons}, R.~C., {et~al.} 2023, \apj, 945, 35, \dodoi{10.3847/1538-4357/acba8a}

\bibitem[{{Umeda, H. D.} \& {Nomoto}(2002)}]{2002ApJ...565..385U}
{Umeda, H. D.}, \& {Nomoto}, K. 2002, \apj, 565, 385, \dodoi{10.1086/323946}

\bibitem[{{Umeda, H. D.} \& {Nomoto}(2003)}]{2003Natur.422..871U}
---. 2003, \nat, 422, 871, \dodoi{10.1038/nature01571}

\bibitem[{{Umeda, H. D.} {et~al.}(2002){Umeda, H. D.}, {Nomoto}, {Tsuru}, \& {Matsumoto}}]{2002ApJ...578..855U}
{Umeda, H. D.}, {Nomoto}, K., {Tsuru}, T.~G., \& {Matsumoto}, H. 2002, \apj, 578, 855, \dodoi{10.1086/342650}

\bibitem[{{Umeda, H. R.} {et~al.}(2022){Umeda, H. R.}, {Ouchi}, {Nakajima}, {Isobe}, {Aoyama}, {Harikane}, {Ono}, \& {Matsumoto}}]{2022ApJ...930...37U}
{Umeda, H. R.}, {Ouchi}, M., {Nakajima}, K., {et~al.} 2022, \apj, 930, 37, \dodoi{10.3847/1538-4357/ac602d}

\bibitem[{{Umeda, H.D.} \& {Nomoto}(2008)}]{2008ApJ...673.1014U}
{Umeda, H.D.}, \& {Nomoto}, K. 2008, \apj, 673, 1014, \dodoi{10.1086/524767}

\bibitem[{{Vincenzo} {et~al.}(2016){Vincenzo}, {Belfiore}, {Maiolino}, {Matteucci}, \& {Ventura}}]{2016MNRAS.458.3466V}
{Vincenzo}, F., {Belfiore}, F., {Maiolino}, R., {Matteucci}, F., \& {Ventura}, P. 2016, \mnras, 458, 3466, \dodoi{10.1093/mnras/stw532}

\bibitem[{{Virtanen} {et~al.}(2020){Virtanen}, {Gommers}, {Oliphant}, {Haberland}, {Reddy}, {Cournapeau}, {Burovski}, {Peterson}, {Weckesser}, {Bright}, {van der Walt}, {Brett}, {Wilson}, {Millman}, {Mayorov}, {Nelson}, {Jones}, {Kern}, {Larson}, {Carey}, {Polat}, {Feng}, {Moore}, {VanderPlas}, {Laxalde}, {Perktold}, {Cimrman}, {Henriksen}, {Quintero}, {Harris}, {Archibald}, {Ribeiro}, {Pedregosa}, {van Mulbregt}, \& {SciPy 1. 0 Contributors}}]{2020NatMe..17..261V}
{Virtanen}, P., {Gommers}, R., {Oliphant}, T.~E., {et~al.} 2020, Nature Methods, 17, 261, \dodoi{10.1038/s41592-019-0686-2}

\bibitem[{{Webbink}(1984)}]{1984ApJ...277..355W}
{Webbink}, R.~F. 1984, \apj, 277, 355, \dodoi{10.1086/161701}

\bibitem[{{Wise} {et~al.}(2012){Wise}, {Turk}, {Norman}, \& {Abel}}]{2012ApJ...745...50W}
{Wise}, J.~H., {Turk}, M.~J., {Norman}, M.~L., \& {Abel}, T. 2012, \apj, 745, 50, \dodoi{10.1088/0004-637X/745/1/50}

\bibitem[{{Woosley} {et~al.}(2002){Woosley}, {Heger}, \& {Weaver}}]{2002RvMP...74.1015W}
{Woosley}, S.~E., {Heger}, A., \& {Weaver}, T.~A. 2002, Reviews of Modern Physics, 74, 1015, \dodoi{10.1103/RevModPhys.74.1015}

\bibitem[{{Xu} {et~al.}(2022){Xu}, {Ouchi}, {Rauch}, {Nakajima}, {Harikane}, {Sugahara}, {Komiyama}, {Kusakabe}, {Fujimoto}, {Isobe}, {Kim}, {Ono}, \& {Zahedy}}]{2022ApJ...929..134X}
{Xu}, Y., {Ouchi}, M., {Rauch}, M., {et~al.} 2022, \apj, 929, 134, \dodoi{10.3847/1538-4357/ac5e32}

\bibitem[{{Yajima} {et~al.}(2022){Yajima}, {Abe}, {Khochfar}, {Nagamine}, {Inoue}, {Kodama}, {Arata}, {Dalla Vecchia}, {Fukushima}, {Hashimoto}, {Kashikawa}, {Kubo}, {Li}, {Matsuda}, {Mawatari}, {Ouchi}, \& {Umehata}}]{2022MNRAS.509.4037Y}
{Yajima}, H., {Abe}, M., {Khochfar}, S., {et~al.} 2022, \mnras, 509, 4037, \dodoi{10.1093/mnras/stab3092}

\bibitem[{{Zhang}(1996)}]{atom_fe_Z}
{Zhang}, H. 1996, \aaps, 119, 523

\end{thebibliography}
\bibliographystyle{aasjournal}

\end{document}